\documentclass[aps,prx,twocolumn,superscriptaddress,longbibliography]{revtex4-2}
\usepackage{amsmath}
\usepackage{graphicx}
\usepackage{amsfonts}
\usepackage{amssymb}
\usepackage{dsfont}
\usepackage{amsmath} % or simply amstext

\usepackage{hyperref}

\newcommand{\hata}{\hat{a}}
\newcommand{\Omegazeroe}{\Omega_{\mathrm{0e}}}

\newcommand{\Omegazeroetilde}{\tilde{\Omega}_{\mathrm{0e}}}
\newcommand{\Omegarms}{\tilde{\Omega}_{\mathrm{RMS}}}

\newcommand{\Omegaoneetilde}{\tilde{\Omega}_{\mathrm{1e}}}
\newcommand{\Omegaonee}{\Omega_{\mathrm{1e}}}
\newcommand{\Omegaae}{\Omega_{\mathrm{ae}}}
\newcommand{\Omegaaetilde}{\tilde{\Omega}_{\mathrm{ae}}}

\newcommand{\nkl}{n_{kl}}

\newcommand{\Vje}{\tilde{V}_{j\mathrm{e}}}

\newcommand{\Vjedot}{\partial_{t'}\tilde{V}_{j\mathrm{e}}}

\newcommand{\Deltajkrpm}{\Delta_{kl,\pm}^{\mathrm{cav}}}

\newcommand{\Deltakljpm}{\Delta_{kl,\pm}^{(j)}}

\newcommand{\HJC}{\hat{H}_{\mathrm{JC}}}
\newcommand{\tramp}{t_{\mathrm{ramp}}}

\newcommand{\hatsigma}{\hat{\boldsymbol{\sigma}}}
\newcommand{\rma}{\mathrm{a}}
\newcommand{\rme}{\mathrm{e}}
\newcommand{\rmd}{\mathrm{d}}

\newcommand{\rmbpm}{\mathrm{b}_{\pm}}

\newcommand{\Herr}{\hat{H}_{\mathrm{err}}}

\newcommand{\Hzero}{\hat{H}_0}

\newcommand{\Hstirap}{\hat{H}_{\mathrm{0}}}

\newcommand{\ket}[1]{|#1\rangle}
\newcommand{\bra}[1]{\langle#1|}

\newcommand{\Deltakl}{\Delta_{kl,\sigma}^{(j)}}

\newcommand{\VRMS}{\tilde{V}_{\mathrm{RMS}}}
\newcommand{\Tonecav}{T_{1,\mathrm{cav}}}
\newcommand{\Ttwocav}{T_{2,\mathrm{cav}}}

\newcommand{\ncav}{\bar{n}_{\mathrm{cav}}}
\newcommand{\omegar}{\omega_{\mathrm{cav}}}

\newcommand{\LQF}{\mathcal{L}^{\mathrm{QF}}}
\newcommand{\Toeo}{(T_{1})_{\mathrm{e}1}}
\newcommand{\Qdiel}{Q_{\mathrm{diel}}}
\hypersetup{colorlinks = true, urlcolor = blue, linkcolor = blue, citecolor = blue}

\begin{document}

%%%%%%%%%%%%%%%%%%%%%%%%%%%%
\title{Analytic Design of Accelerated Adiabatic Gates in Realistic Qubits:  General Theory and Applications to Superconducting Circuits}

\author{F. Setiawan}\email{setiawan@uchicago.edu}
\affiliation{Pritzker School of Molecular Engineering, University of Chicago, 5640 South Ellis Avenue, Chicago, Illinois 60637, USA}
\author{Peter Groszkowski}
\affiliation{Pritzker School of Molecular Engineering, University of Chicago, 5640 South Ellis Avenue, Chicago, Illinois 60637, USA}
\author{Hugo Ribeiro}
\affiliation{Max Planck Institute for the Science of Light, Staudtstra{\ss}e 2, 91058 Erlangen, Germany}
\author{Aashish A. Clerk}
\affiliation{Pritzker School of Molecular Engineering, University of Chicago, 5640 South Ellis Avenue, Chicago, Illinois 60637, USA}
\date{\today}

\begin{abstract}
Shortcuts to adiabaticity is a general method for speeding up adiabatic quantum protocols, and has many potential applications in quantum information processing.  Unfortunately, analytically constructing shortcuts to adiabaticity for systems having complex interactions and more than a few levels is a challenging task.  This is usually overcome by assuming an idealized Hamiltonian 
[e.g.,~only a limited subset of energy levels are retained, and the rotating-wave approximation (RWA) is made]. Here we develop an \textit{analytic} approach that allows one to go beyond these limitations. Our method is general and results in analytically derived pulse shapes that correct both nonadiabatic errors as well as non-RWA errors. We also show that our approach can yield pulses requiring a smaller driving power than conventional nonadiabatic protocols. We show in detail how our ideas can be used to analytically design high-fidelity single-qubit ``tripod" gates in a realistic superconducting fluxonium qubit.

\end{abstract}

\maketitle
\section{Introduction}

\begin{figure}[t!]
\centering
\includegraphics[width=\linewidth]{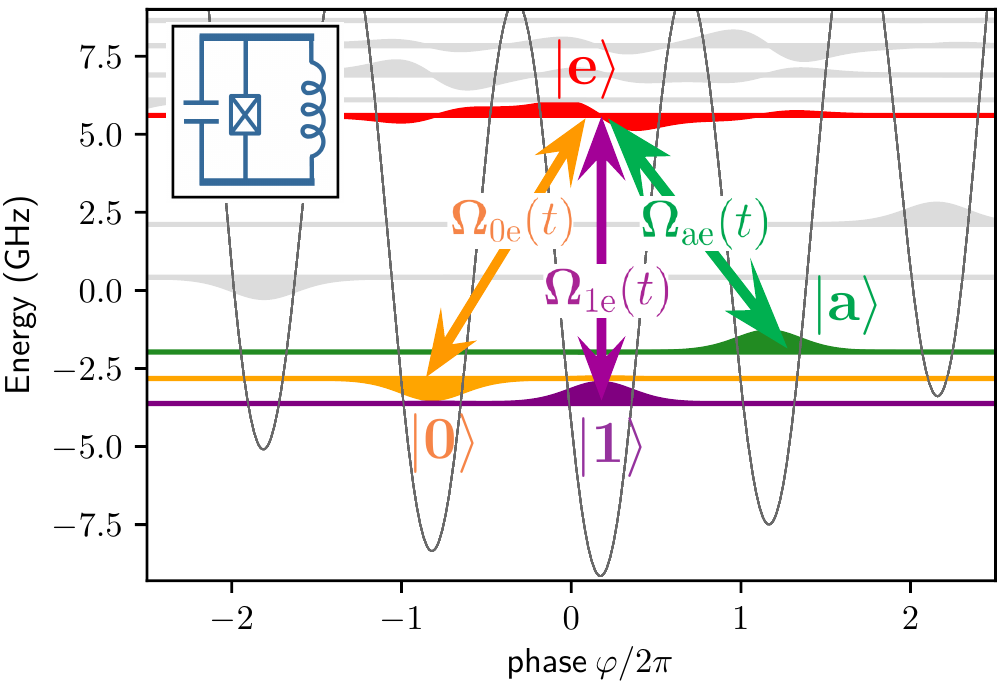}
\caption{A single-qubit gate realized using a four-level tripod energy structure consisting of three lower energy levels ($|0\rangle$, $|1\rangle$ and $|\text{a}\rangle$) resonantly coupled to an excited state $|\text{e}\rangle$ by three different driving tones: with envelopes $\Omega_{0\text{e}}(t)$, $\Omega_{1\text{e}}(t)$, and  $\Omega_{\text{a}\text{e}}(t)$, respectively. This tripod-energy structure can be engineered in a multienergy spectrum (shown above are the ten lowest energy levels) of a fluxonium qubit, whose device schematic is shown in the inset. The driving fields can drive unwanted transitions that give rise to coherent errors, e.g., transitions between the energy levels outside the tripod structure (shown using light gray) and the energy levels in the tripod structure as well as the transitions inside the tripod structure driven by nonresonant tones.}\label{fig:tripodfluxonium}
\end{figure}

Quantum gates based on adiabatic evolution~\cite{Zanardi1999,Pachos1999,duan2001geometric,Kis2002Qubit,Faoro2003Non,Solinas2003Holonomic,Moller2008Geometric,frees2019adiabatic} are generally desirable because of their intrinsic robustness against imperfections in control pulses, and have been implemented in a variety of platforms (see e.g., Refs.~\cite{Wu2013Geometric,Toyoda2013Realization,Huang2019Experimental}). They, however, require extremely long evolution times, making them potentially susceptible to dissipation and noise.  An intriguing possibility is to try to ``accelerate" adiabatic gates using techniques drawn from the field of shortcuts to adiabaticity (STA)~\cite{demirplak2003adiabatic,demirplak2005assisted,berry2009transitionless,Ibanez2012Multiple,Guery2019Shortcuts}.  STA protocols seek to modify pulses to completely cancel nonadiabatic errors.  They are usually developed for simple evolutions that do not correspond to a true quantum gate, as their form is tied to a specific choice of initial state.  However, they can be adapted for true gates.   
Ribeiro and Clerk.~\cite{Ribeiro2019Accelerated} showed how a particular shortcut approach [superadiabatic transitionless driving (SATD) \cite{Baksic2016Speeding}] could be used to accelerate a true (arbitrary) single-qubit gate, the paradigmatic ``tripod" adiabatic gate
introduced in Refs.~\cite{duan2001geometric,Kis2002Qubit} (see Fig.~\ref{fig:tripodfluxonium}).  Other STA approaches to gates were presented in Refs.~\cite{zhang2015fast,song2016shortcuts,wang2018experimental,Yan2019Experimental,santos2020optimizing}. Note that STA-accelerated gates are conceptually distinct from the so-called nonadiabatic holonomic gates (see e.g., Ref.~\cite{sjoqvist2012non}) and have distinct advantages \cite{Ribeiro2019Accelerated}. 

While the above results are promising, they are limited by a constraint that plagues most STA approaches to quantum control:  accelerated protocols can  be derived analytically for only few-level systems (or systems that reduce to uncoupled few-level systems).  Furthermore, one needs to ignore fast-oscillating nonresonant terms [i.e., one must necessarily make the rotating-wave approximation (RWA)].  In many realistic settings, unwanted non-RWA dynamics and couplings to higher levels cannot be ignored, and will limit the operation fidelity even if nonadiabatic errors are suppressed.  For this reason, the utility of analytic STA protocols for high-fidelity operations in  complex systems has remained unclear.

%%%%%%%%%%%%%%%%%%%%%%%%%%

In this paper, we  present a generic approach for improving STA protocols in settings where assuming an idealized dynamics is not possible (e.g., non-RWA terms cannot be ignored).  The result is a general  method for \textit{analytically} deriving pulse sequences that both fully cancel nonadiabatic errors, and partially mitigate non-RWA errors.  To highlight the efficacy of our approach, we focus on a specific, experimentally relevant setting:  an accelerated geometric tripod gate implemented in a fluxonium-style 
superconducting circuit~\cite{manucharyan2009fluxonium,Earnest2018Realization,Nguyen2019High,Helin2021Universal}. 
The isolated qubit levels of this system make conventional approaches to gates problematic, providing motivation for new ideas that do not require a direct coupling of qubit levels.  The so-called tripod gate \cite{duan2001geometric,Kis2002Qubit}
is a natural candidate.  However, as shown in Fig.~\ref{fig:tripodfluxonium}, a fluxonium circuit has a complex level structure, implying that nonresonant, non-RWA corrections will be important. Using realistic parameters compatible with experiment, we show (via full master-equation solutions including dissipation,  non-RWA effects, and power constraints from cavity-based driving) that, subject to realistic noise, our
 enhanced STA,  gate achieves gate fidelities of $0.9991$-$0.9997$ with a gate time of $t_g = 100$ ns. 

Our gate performance is roughly comparable to results obtained using fully numerical optimal control on a related superconducting circuit \cite{Mohamed2020Universal}.
It also demonstrates that the accelerated tripod gate can be advantageous, despite the ability to realize a perfectly isolated tripod level structure.
 Note that the use of our enhanced protocol is crucial:  if one simply uses the STA derived without corrections, the fidelity error for the same gate time is orders of magnitude higher.

The paper is organized as follows. We begin in Sec.~\ref{sec:STAmulti} by introducing the most general version of our problem:  how can one analytically design STA protocols in complex multilevel systems?   In Sec.~\ref{sec:quantumgate}, we briefly review the basic (RWA)  geometric tripod gate \cite{duan2001geometric,Kis2002Qubit} as well as its accelerated version \cite{Ribeiro2019Accelerated}.  In Sec.~\ref{sec:chirp}, we go beyond the RWA, and discuss how in general STA approaches can be further enhanced to mitigate nonresonant errors.  In Sec.~\ref{sec:fluxonium}, we explore the utility of these methods by applying them to design an accelerated gate in a realistic fluxonium superconducting circuit.  Results for gate performance are presented in Sec.~\ref{sec:gateperformance}, and comparison against a simpler ``direct-driving" gate is presented in Sec.~\ref{sec:directdriving}.  We summarize our results in Sec.~\ref{sec:conclusions}. 

\section{General problem: Shortcut-to-adiabaticity approaches for complex driven systems}\label{sec:STAmulti}
We begin by considering a generic driven multilevel system whose Hamiltonian in the laboratory frame has the form
\begin{equation}\label{eq:H}
    \hat{H}(t) = \sum_{k}\varepsilon_k |k\rangle \langle k| + \left(V(t) \sum_{ \substack{(k,l) \, \rvert \\ l>k}}n_{kl} |k\rangle \langle l|+ \mathrm{H.c.}\right),
\end{equation}
where $\varepsilon_{k}$ and $|k\rangle$ are the eigenenergies and eigenstates of the undriven system, respectively, and $n_{kl} = \langle k| \hat{n}|l \rangle$ is an effective dipole matrix element. The full control pulse $V(t)$ consists of several distinct drive tones $\omega_j$, each associated with a slowly varying complex envelope $V_j(t)$, i.e.,
\begin{equation}\label{eq:V}
    V(t) = \frac{1}{2}\sum_{j} \left( V_{j}(t) e^{i \omega_{j}t} + \textrm{H.c.} \right).
\end{equation}

We next move to an interaction picture defined by $\hat{U}_{\mathrm{diag}} = e^{-i\hat{H}_{\mathrm{diag}}t}$, where $\hat{H}_{\mathrm{diag}} = \sum_{k} \varepsilon_k |k\rangle \langle k|$.  The Hamiltonian in this frame takes the general form
 \begin{equation}\label{eq:Htot}
\hat{H}(t) = \Hstirap(t) + \Herr(t).
\end{equation}
Here $\Hstirap(t)$ describes resonant processes, and is time dependent only through its dependence on the envelope functions $V_j(t)$.  Defining $\varepsilon_{kl} = \varepsilon_l - \varepsilon_k$, we have
\begin{equation}
    \Hstirap(t) = \frac{1}{2}\sum_{j} \sum_{ \substack{(k,l) \, \rvert \\ l>k \, \& \, \varepsilon_{kl} = \omega_j}}
    \Big(V_j(t) n_{kl}|k\rangle \langle l|+\mathrm{H.c.} 
    \Big).
\end{equation}
In contrast, $\Herr(t)$ describes all nonresonant processes
\begin{align}\label{eq:HerrIntro}
    \Herr(t) & =  \frac{1}{2}\Bigg( \sum_{j} 
         \sum_{\substack{(k,l) \, \rvert \\ l>k\,\&\, \varepsilon_{kl} \neq \omega_j}}
    V_j(t) e^{-i(\varepsilon_{kl}-\omega_j)t} n_{kl}|k\rangle \langle l|
\nonumber\\
& +
     \sum_j \sum_{\substack{(k,l) \, \rvert \\ l>k\,\&\,\varepsilon_{kl} \neq -\omega_j}}
    V^*_j(t) e^{-i(\varepsilon_{kl}+\omega_j)t} n_{kl}|k\rangle \langle l| \Bigg) +\mathrm{H.c.} 
\end{align}

The standard next step in most analytic STA approaches to quantum control is to make the RWA:  one assumes that the energy detunings in $\Herr(t)$ are sufficiently large that this error Hamiltonian can be approximated as zero.  The result is a much simpler Hamiltonian that involves only the slowly varying amplitudes $V_j(t)$ and which typically couples only a small subset of levels.  It is in this context that many exact STA protocols have been derived; these protocols yield a perfect, error-free evolution within the RWA.  Examples range from accelerated versions of the two-level Landau-Zener problem~\cite{demirplak2003adiabatic,demirplak2008consistency,berry2009transitionless,Ibanez2012Multiple}, to more complex three-level~\cite{Baksic2016Speeding,huang2016fast,kang2016fast,Song2016Physically,wu2017superadiabatic} and four-level~\cite{Ribeiro2019Accelerated} protocols.

Despite the power of the above STA approaches, they address only nonadiabatic errors associated with the RWA Hamiltonian $\Hstirap(t)$.  A crucial question is whether they can also be adapted to address additional errors arising from the non-RWA dynamics described by $\Herr(t)$.  Corrections to the RWA are important in many physical systems if one is interested in realizing truly high-fidelity operations.  One could in principle try to derive an STA for the full multilevel fast-oscillating Hamiltonian $H(t)$ but in most cases this is completely infeasible.  Not only does this involve dealing with a large-dimensional Hilbert space but it also involves working with a starting Hamiltonian that has extremely fast time-dependent terms [i.e., in $\Herr(t)$], and hence is nowhere close to an adiabatic limit.

A central goal of this work is to present a much more tractable approach to adapting exact STA protocols so that they also mitigate nonresonant, non-RWA errors.  Our method is ultimately perturbative, and amounts to modifying the original (RWA) STA protocol to correct the leading effects of $\Herr(t)$.  We stress that our approach retains the crucial feature of the original STA pulse sequence of being described and derived fully analytically (i.e., no recourse is made to numerical optimal-control approaches).  While our method is extremely general, we focus in what follows on a particularly promising protocol involving accelerated geometric gates using a tripod level structure \cite{duan2001geometric,Kis2002Qubit} [see Fig.~\ref{fig:tripod}(a)].
The general approach we present 
is completely distinct
from recent work~\cite{Boyers2019Floquet,Claeys2019Floquet} which deliberately introduce additional high-frequency oscillatory terms to a RWA Hamiltonian to approximately engineer desired STA protocols.

\section{Review: Accelerated adiabatic quantum gates}~\label{sec:quantumgate}

In this section we briefly review the basic geometric tripod gate introduced in Refs.~\cite{duan2001geometric,Kis2002Qubit,Moller2008Geometric,Liu2017Superadiabatic} and its accelerated version \cite{Ribeiro2019Accelerated}.  All these analyses were done in the context of a simplified four-level RWA Hamiltonian.  Our review here sets the stage for our following discussion on how these approaches can be modified and effectively implemented in a realistic multilevel superconducting circuit where non-RWA effects play a crucial role. 

\subsection{Double Stimulated Raman Adiabatic Passage (STIRAP) protocol in an ideal tripod system}
\begin{figure}[t!]
\centering
\includegraphics[width=\linewidth]{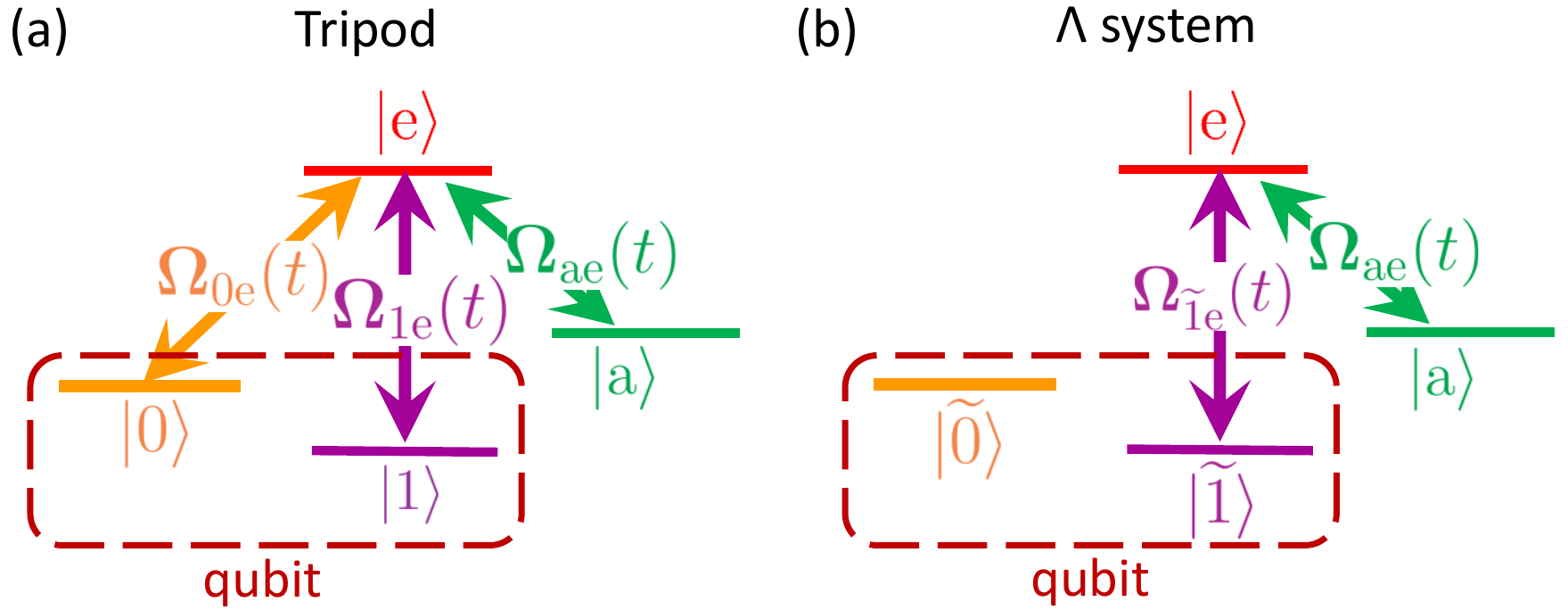}
\caption{(a) An ideal tripod system consisting of three lower energy levels (the qubit states $|0\rangle$ and $|1\rangle$ as well as the auxiliary state $|\text{a}\rangle$) resonantly coupled to an excited state $|\text{e}\rangle$ by three different driving tones, with envelopes $\Omega_{0\text{e}}(t)$, $\Omega_{1\text{e}}(t)$, and $\Omega_{\text{a}\text{e}}(t)$, respectively. (b) An effective $\Lambda$ system (consisting of the state $|\tilde{1}\rangle$ and the state $|\mathrm{a}\rangle$ resonantly coupled to the state $|\mathrm{e}\rangle$) used to describe the dynamics of the tripod system. The control pulses $\Omega_{\tilde{1}\mathrm{e}}(t)$ and $\Omega_{\mathrm{ae}}(t)$ perform a double stimulated Raman adiabatic passage (STIRAP) protocol that cyclically evolves the zero-energy dark states in the $\Lambda$ system.}\label{fig:tripod}
\end{figure}

Starting with the full driven Hamiltonian in Eq.~\eqref{eq:H}, we assume a situation where within the RWA we realize a so-called tripod level configuration [see Fig.~\ref{fig:tripod}(a)].
An ideal tripod system consists of three lower levels (labeled by $|0\rangle, |1\rangle$ and $|\text{a}\rangle$) that are controllably coupled to a common excited state $|\text{e}\rangle$ [see Fig.~\ref{fig:tripod}(a)].  Denoting these (complex) couplings as 
$\Omega_{j\text{e}}(t)$ ($j=0,1,\mathrm{a}$), we can write the  tripod Hamiltonian as
($\hbar = 1$)
\begin{align}\label{eq:Hstirap}
    \Hstirap(t) &= \frac{1}{2} \left[\Omegazeroe(t) |0\rangle \langle \rme| + \Omegaonee(t)|1\rangle \langle \rme| + \Omegaae(t) |\rma\rangle \langle \rme| + \mathrm{H.c.} \right],
\end{align}
where  
\begin{align}\label{eq:Vme}
\Omega_{j\mathrm{e}}(t)  &= V_{j\mathrm{e}}(t)n_{j\mathrm{e}}
\end{align}
for $j = 0,1,\rma$.
We take the states $|0\rangle$ and $|1\rangle$ to encode a logical qubit, while $|\mathrm{a}\rangle$ and $|\mathrm{e}\rangle$ serve as  auxiliary states used to perform gate operations.

The basic idea of the gate is that $\Hstirap(t)$ always has two degenerate zero-energy adiabatic eigenstates, and hence cyclic adiabatic evolution can result in a nontrivial geometric $2 \times 2$ unitary in this subspace.  To understand this more concretely, we follow Ref.~\cite{Ribeiro2019Accelerated}, and consider control pulses of the form
\begin{subequations}\label{eq:omegadrive}
\begin{align}
\Omegazeroe(t) &= \Omega_0 \cos(\alpha) \sin [\theta(t)], \\
\Omegaonee(t) &= \Omega_0 \sin(\alpha) \sin [\theta(t)]e^{i\beta},\\
\Omegaae(t) &= \Omega_0 \cos[\theta(t)]e^{i\gamma(t)}\label{eq:omegaae}.
\end{align} 
\end{subequations}
The angles $\alpha$ and $\theta$ control the relative magnitudes of the pulses, while $\beta$ and $\gamma$ control relative phases; $\alpha$ and $\beta$ are time independent and we only require $\theta(t)$ and $\gamma(t)$ to be time dependent.  
The overall amplitude $\Omega_0$ sets the instantaneous adiabatic gap of $\Hstirap(t)$, which we choose to keep constant:
\begin{equation}
    \Omega_{\mathrm{ad}}(t) \equiv \frac{1}{2}\sqrt{|\Omega_{0\rme}(t)|^2 + |\Omega_{1\rme}(t)|^2 + |\Omega_{\rma\rme}(t)|^2} = \frac{\Omega_0}{2}.
    \label{eq:AdiabaticGap}
\end{equation}
At every instant in time, $\Hstirap(t)$ has two zero-energy dark states (orthogonal to $|\rme\rangle$), and bright states at energy $\pm \Omega_0/2$. 

The tripod system in Fig.~\ref{fig:tripod}(a) can be related to an effective three-level $\Lambda$ system by moving into a suitable frame defined by the time-independent control pulse parameters $\alpha$ and $\beta$~\cite{Bergmann1998Coherent,Vitanov2017Stimulated} [Fig.~\ref{fig:tripod}(b)]. In this representation, the gate corresponds to a ``double STIRAP protocol" in the $\Lambda$ system, where the dark state 
\begin{equation}\label{eq:d2}
|\rmd(t)\rangle = \cos[\theta(t)]|\tilde{1}\rangle - e^{i\gamma(t)}\sin[\theta(t)]|\rma\rangle
\end{equation}
undergoes a cyclic adiabatic evolution $|\tilde{1}\rangle \rightarrow |\rma\rangle \rightarrow |\tilde{1}\rangle$.  This requires an appropriate cyclic variation of the pulse  parameter $\theta(t)$ ~(see Ref.~\cite{Ribeiro2019Accelerated} and Appendix~\ref{sec:tripodgate} for details).  This cyclic evolution can result in a Berry phase.  We take the gate to start at $t=0$ and end at $t = t_g$.   For the case where the pulse parameter $\gamma(t)$ is chosen as
\begin{equation}\label{eq:gamma}
\gamma(t) = \gamma_0 \Theta \left(t - \frac{t_g}{2} \right),
\end{equation}
with $\Theta(t)$ being the Heaviside step function, this geometric phase is simply $\gamma_0$~\cite{Ribeiro2019Accelerated}.

In the adiabatic limit $\dot{\theta}(t)/\Omega_0 \rightarrow 0$, the gate unitary in the qubit subspace is given by~\cite{Ribeiro2019Accelerated}
\begin{subequations}\label{eq:UQubitAdiabatic}
\begin{align}
    \hat{U}_{\mathrm{G},01} & = 
    \exp\left(-i \gamma_0 /2 \right)
    \exp \left( -i\frac{\gamma_0}{2}\boldsymbol {n}\cdot \hatsigma_{01} \right),
    \\
    \boldsymbol{n} &= [\sin (2\alpha) \cos (\beta), \sin(2\alpha)\sin(\beta),\cos(2\alpha)],
\end{align}
\end{subequations}
where $\hatsigma_{01} = (|0\rangle \langle 1| + \mathrm{H.c.}, -i |0\rangle \langle 1| + \mathrm{H.c.}, |0\rangle \langle 0| - |1\rangle\langle 1|)$ is the vector of the Pauli matrices in the qubit subspace. For example, the X gate can be realized by using the angle parameters $\alpha = \pi/4$, $\beta =0$, and $\gamma_0 = \pi$.

\subsection{Accelerated tripod gates}\label{sec:accelerated}
The geometric tripod gate yields a perfect gate fidelity in the adiabatic limit where the protocol time is infinitely longer than $1/\Omega_0$.  In many realistic systems, dissipative effects involving the lower tripod levels make such long evolution times infeasible.  It would thus be desirable to reduce the gate time without introducing nonadiabatic errors.  This is exactly the goal of STA methods~\cite{demirplak2003adiabatic,demirplak2005assisted,berry2009transitionless,Baksic2016Speeding,Ribeiro2019Accelerated,demirplak2008consistency}.

Following Ref.~\cite{Ribeiro2019Accelerated}, we consider an STA protocol based on the SATD method \cite{Baksic2016Speeding}, where nonadiabatic errors are mitigated by having the system follow a dressed adiabatic eigenstate (see Appendix~\ref{sec:SATD}).  The accelerated protocol is implemented by simply modifying the complex envelope of the original control pulse~\cite{Baksic2016Speeding,Ribeiro2019Accelerated,Zhou2017Accelerated,roque2020engineering}. Specifically, the SATD protocol requires that one corrects the original pulses via
\begin{subequations}\label{eq:omegaSATD}
\begin{align}
\Omegazeroe(t) &\rightarrow \Omegazeroetilde(t) \equiv\Omega_0 \cos(\alpha)\left[\sin[\theta(t)] + 4 \frac{\cos[\theta(t)]\ddot{\theta}(t)}{\Omega_0^2 + 4 \dot{\theta}^2(t)}\right],\\
\Omegaonee(t) &\rightarrow \Omegaoneetilde(t) \equiv \Omega_0 \sin(\alpha)e^{i\beta}\left[\sin[\theta(t)] + 4 \frac{\cos[\theta(t)]\ddot{\theta}(t)}{\Omega_0^2 + 4 \dot{\theta}^2(t)}\right],\\
\Omegaae(t) &\rightarrow \Omegaaetilde(t) \equiv \Omega_0 e^{i\gamma(t)}\left[\cos[\theta(t)] - 4 \frac{\sin[\theta(t)]\ddot{\theta}(t)}{\Omega_0^2 + 4 \dot{\theta}^2(t)}\right],
\end{align}
\end{subequations}
where the angle $\gamma(t)$ [Eq.~\eqref{eq:gamma}] remains unchanged.  It can be shown~\cite{Ribeiro2019Accelerated} that the resulting accelerated
protocol obtained using correction in Eq.~\eqref{eq:omegaSATD} achieves the same unitary $\hat{U}_{\mathrm{G},01}$ in the qubit subspace as in the adiabatic limit [see~Eq.~\eqref{eq:UQubitAdiabatic}]. 
In what follows, we use a tilde to denote SATD-corrected pulse parameters.

\begin{figure}[t!]
\centering
\includegraphics[width=\linewidth]{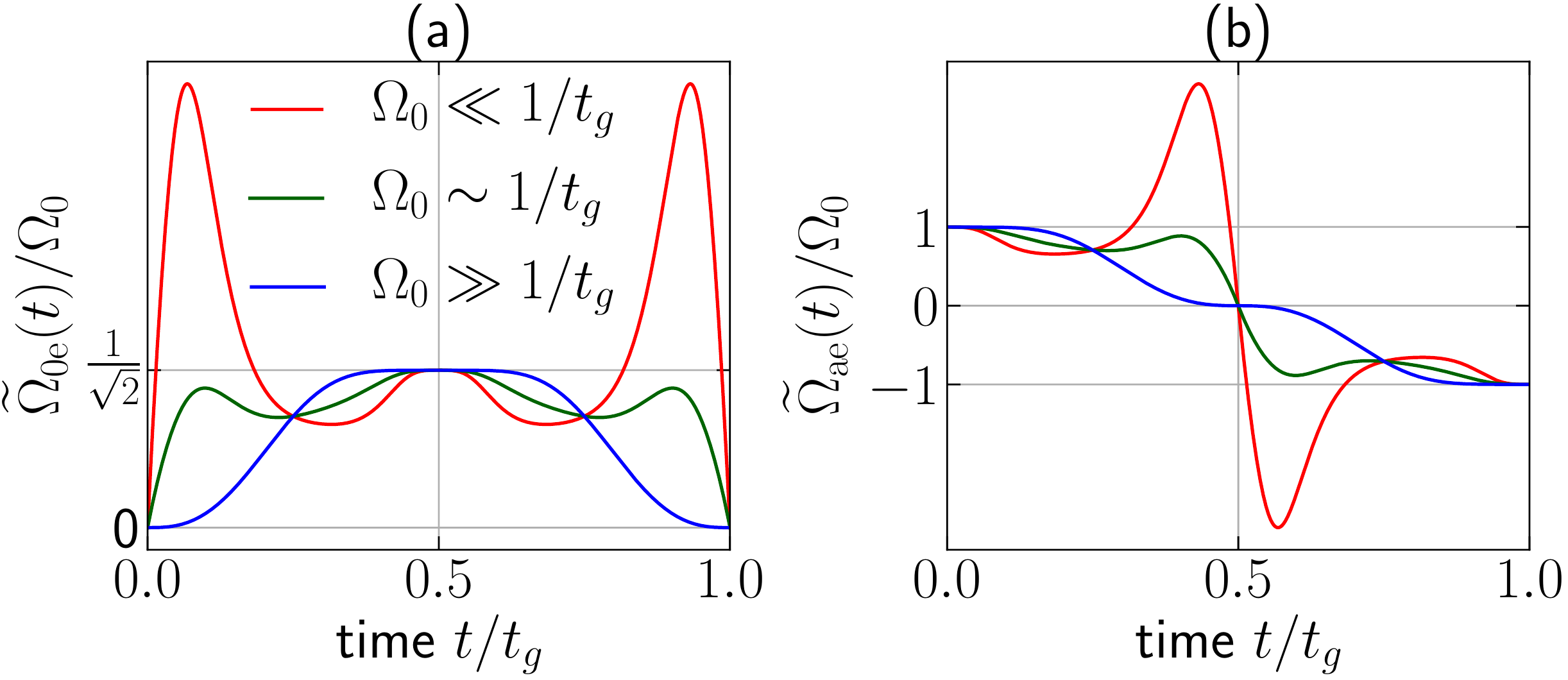}
\caption{Time profiles of the envelopes of the three driving tones used to realize X gates in an ideal tripod system: (a) $\tilde{\Omega}_{0\mathrm{e}}(t) = \tilde{\Omega}_{1\mathrm{e}}(t)$, (b) $\tilde{\Omega}_{\mathrm{ae}}(t)$. Pulses are calculated with use of Eq.~\eqref{eq:omegaSATD} with $\theta(t)$ given in Eq.~\eqref{eq:theta} and other angle parameters given by $\alpha = \pi/4$, $\beta =0$, and $\gamma_0 = \pi$. Shown here are pulses for a fixed gate time $t_g$ but with different uncorrected gap frequencies: $\Omega_0/2\pi = 4/t_g$ (red curve), $\Omega_0/2\pi = 2/t_g$ (green curve), and $\Omega_0/2\pi = 0.2/t_g$ (blue curve). The SATD correction to the adiabatic pulse becomes larger as $\Omega_0$ decreases.}
\label{fig:omegapulse}
\end{figure} 

\subsection{Infinite family of perfect RWA protocols}
\label{subsec:InfiniteFamily}

For our ideal (RWA) tripod systems, our SATD approach yields an {\it infinite} number of perfect protocols (i.e., pulse sequences) that realize a given gate in a fixed gate time $t_g$. These protocols are indexed by $\Omega_0$ (see~Eq.~(\ref{eq:AdiabaticGap})), which is the scale of amplitudes of the {\it uncorrected} pulse (and the corresponding time-independent adiabatic gap).  For every choice of $\Omega_0$, there is a corresponding SATD protocol [given by Eq.~\eqref{eq:omegaSATD}] that yields a pulse sequence with a perfect gate fidelity.  At a heuristic level, for $\Omega_0 \gg 1/t_g$ the uncorrected protocol is already almost in the adiabatic limit, meaning that the additional SATD modification of pulses will be minimal.  In contrast, for $\Omega_0 < 1/t_g$, the SATD correction to the original pulse shape will be extremely large ( to cancel nonadiabatic errors).  Figure~\ref{fig:omegapulse} shows the time profiles of pulse envelopes used to realize perfect X gates in an ideal tripod system for different choices of $\Omega_0 t_g$. Besides the degeneracy in choosing $\Omega_0$, there is also a degeneracy resulting from different choices of the pulse-shape function  $P(t/t_g)$~[Eq.~\eqref{eq:Px}] that determines  $\theta(t)$ [Eq.~\eqref{eq:theta}].

\section{Enhancing STA protocols to mitigate nonresonant errors}\label{sec:chirp}
We now return to the central question of this work:  can STA approaches still be effective in settings where nonresonant, non-RWA processes also degrade fidelity? The non-RWA terms can in general induce energy shifts of the main tripod levels, cause leakage to nontripod levels, and also drive additional higher-order processes; these all represent error mechanisms.  In this section, we present a general strategy for improving STA protocols to partially mitigate non-RWA errors.  For concreteness, we do this in the specific context of the accelerated tripod gate introduced above.  To achieve a gate in a fixed time $t_g$, our strategy has two basic steps:
\begin{enumerate}
    \item We first use the degeneracy of perfect STA protocols that exists in the RWA limit (see~Sec.~\ref{subsec:InfiniteFamily}) to pick a protocol that minimizes the ``size'' (appropriately defined) of our control fields. Since the non-RWA errors increase with increasing pulse amplitude, this step mitigates all non-RWA errors.
    
    \item Next, we use a perturbative approach to partially correct non-RWA errors. We focus on correcting the leading-order error mechanism, which is unwanted energy shifts of the computational states (i.e.~Stark shifts).  As we see in Sec.~\ref{sec:leakage}, there is in general a wide range of gate times where leakage (and higher-order processes) does not play a significant role, meaning that this perturbative approach is very beneficial.  To this end, we modify the SATD pulse shape by chirping the control fields to offset frequency shifts arising from non-RWA terms.  The form of the required chirp can be found {\it analytically} using a perturbative approach. 
\end{enumerate}
As we see later, this two-pronged, {\it fully analytic} approach results in a modified set of pulses that yield exceptional gate performance even when non-RWA effects are included. Thus, our correction strategy is well suited for obtaining  high-fidelity gates. We now discuss each step of our general method in more detail, focusing on the specific case of our accelerated tripod gate. 

\begin{figure}[t!]
\centering
\includegraphics[width=\linewidth]{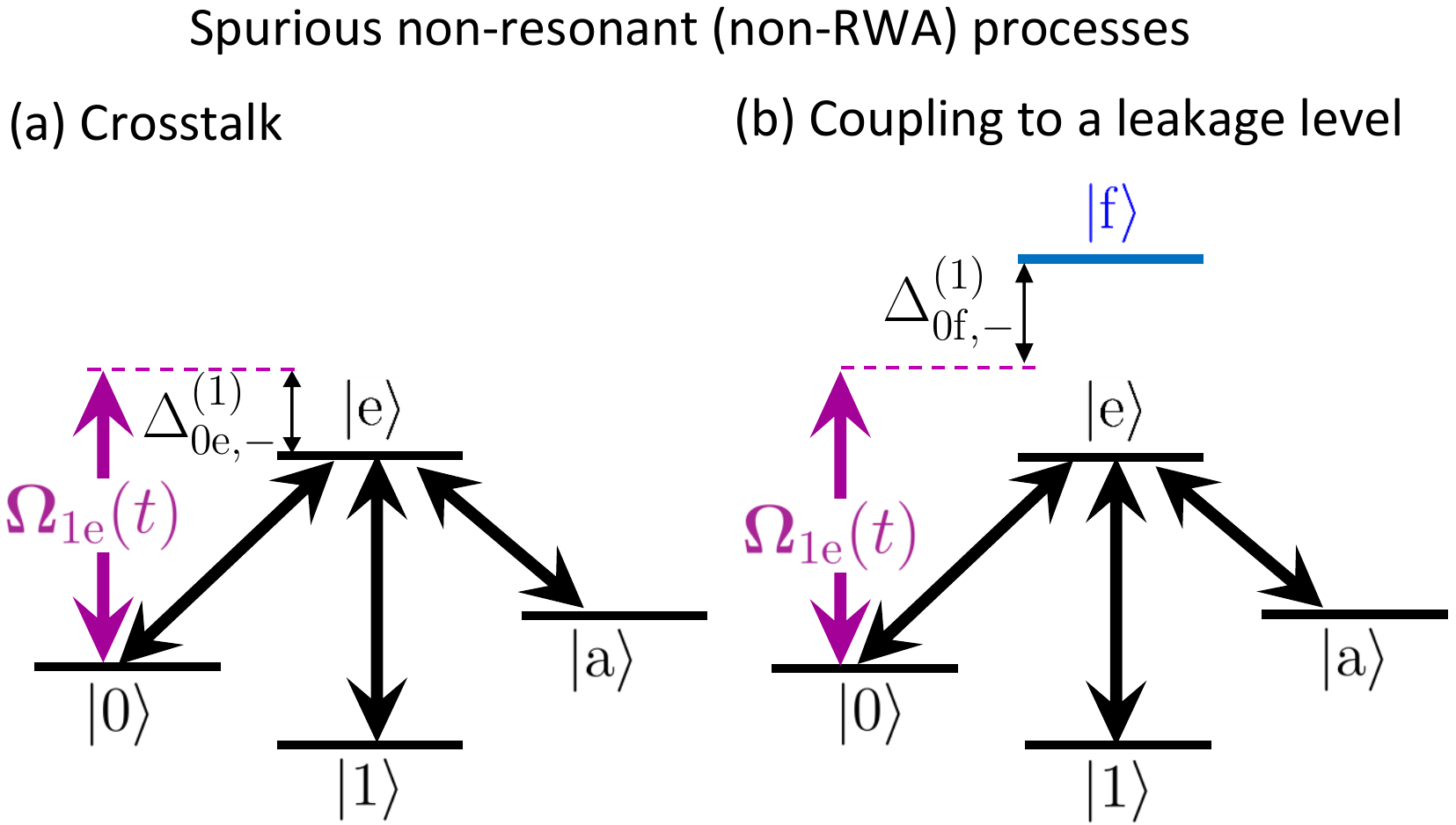}
\caption{Spurious non-RWA processes giving rise to coherent errors. (a) Crosstalk where the driving tone $\Omega_{1\mathrm{e}}(t)$ drives the $|0\rangle \leftrightarrow |\mathrm{e}\rangle$ transition. The frequency of the driving tone $\Omega_{1\mathrm{e}}(t)$ is detuned by $\Delta_{0\mathrm{e},-}^{(1)}$ from the $|0\rangle \leftrightarrow |\mathrm{e}\rangle$ transition.  (b) Coupling to spurious level $|\mathrm{f}\rangle$ where the driving tone $\Omega_{1\mathrm{e}}(t)$ drives the $|0\rangle \leftrightarrow |\mathrm{f}\rangle$ transition. The frequency of the driving tone $\Omega_{1\mathrm{e}}(t)$ is detuned by $\Delta_{0\mathrm{f},-}^{(1)}$ from the $|0\rangle \leftrightarrow |\mathrm{f}\rangle$ transition. }
\label{fig:crosstalk}
\end{figure} 
 
\subsection{Step 1: Power minimization}
\label{subsec:RMSvoltage}

The nonresonant, non-RWA processes described by $\Herr(t)$ [see~Eq.~(\ref{eq:HerrIntro})] yield new unwanted coherent dynamics that will degrade the performance of our gate; example processes are sketched in Fig.~\ref{fig:crosstalk}.  One effect of these terms is to generate effective time-dependent energy shifts of the four levels involved in our tripod gate. 
We define $\Deltakljpm = \varepsilon_l - \varepsilon_k \pm \omega_{j\mathrm{e}}$ as the detuning 
associated with the transition between the energy level $|k\rangle$  and the level $|l\rangle$ associated with the drive tone
$\omega_{j\mathrm{e}}$.  Recall that in our tripod gate, there is a drive tone for each ground-state level (see~Fig.~\ref{fig:tripod}), and hence $j=0,1,\mathrm{a}$.  
Using a Magnus-based approach (see Ref.~\cite{Ribeiro2017Systematic}), one can derive the leading-order, time-dependent energy shift $\delta \varepsilon_k(t)$ of energy level $|k\rangle$ due to $\Herr(t)$.  This energy shift has the usual form expected from second-order perturbation theory, i.e.,
\begin{equation}\label{eq:energyshift}
    \delta\varepsilon_{k}(t) = \sum_{\substack{j = 0,1,\mathrm{a}\\ \sigma = \pm}}     \sum_{l \,  |\Deltakl \neq 0}  \frac{|\Vje(t) n_{kl}|^2}{4\Deltakl}.
\end{equation}
The sum here is over all nonresonant processes that involve the state $|k\rangle$.  We are interested in energy shifts of the four tripod levels (i.e. $k=0,1,\mathrm{a},\mathrm{e}$).  Note that the intermediate states $l$ in 
Eq.~(\ref{eq:energyshift}) includes the four tripod levels [i.e.,~``crosstalk" process; Fig.~\ref{fig:crosstalk}(a)] and nontripod states [i.e.,~couplings to ``leakage" levels; Fig.~\ref{fig:crosstalk}(b)].  
Formally, Eq.~(\ref{eq:energyshift}) is valid in the perturbative limit where $|\Vje(t)\nkl|\ll |\Deltakljpm|$ and the quasistatic limit $|\int_0^{t_g}\Vjedot(t')\nkl e^{-i\Deltakljpm t'} dt'|\ll |\Deltakljpm|$.

The simplest way to mitigate errors associated with the above non-RWA generated energy shifts is to minimize their size by minimizing the SATD-corrected pulse amplitudes $\tilde{V}_{j\mathrm{e}}(t)$.  
We would like to find a simple metric to characterize the size of these amplitudes in a meaningful manner.  
We see that at each instant in time, the relevant quantity is the square of these amplitudes (as the energy shifts are a second-order effect).  This motivates us to characterize the ``size" of our control pulses by the root-mean-square (RMS) voltage of the control field, i.e.,
\begin{equation}\label{eq:VRMStilde}
    \tilde{V}_{\mathrm{RMS}} \equiv \sqrt{\frac{1}{t_g} \int_0^{t_g} \left[\tilde{V}(t)\right]^2 dt}.
\end{equation}
Here $\tilde{V}(t)$ is the total real-valued control pulse function (including the SATD correction), see~Eq.~\eqref{eq:V}. 

In general, coherent errors increase with increasing drive strength (i.e.~pulse amplitude), which we characterize by the RMS voltage of our pulses.
By minimizing $\tilde{V}_{\mathrm{RMS}}$, we can therefore limit the effects of these errors on the performance of our gate. As discussed in Sec.~\ref{sec:powerscale}, there are also other factors that contribute to a desire to limit the overall amplitude of the drive pulses. For example, when the qubit is indirectly driven through a cavity, it is preferable to use a smaller driving power to constrain the cavity photon number. High drive-induced cavity photon numbers are believed to cause heating effects, which may result in a coherence loss of the qubit, potentially further limiting the performance of our gate.

Using the specific form of the SATD pulses in Eqs.~\eqref{eq:omegaSATD}, we can write the RMS voltage in Eq.~\eqref{eq:VRMStilde} as
\begin{align}\label{eq:VRMSSATD}
    \tilde{V}_{\mathrm{RMS}} & \simeq \sqrt{\frac{1}{2t_g} \int_0^{t_g} dt \sum_{j=0,1,\mathrm{a}}\left|\tilde{V}_{j\mathrm{e}}(t)\right|^2} \nonumber\\
&=  \sqrt{\frac{\cos^2\alpha}{|n_{\mathrm{0e}}|^2} + \frac{\sin^2\alpha}{|n_{\mathrm{1e}}|^2} + \frac{1}{|n_{\mathrm{ae}}|^2}} \frac{\Omegarms(t_g)}{2},
\end{align}
where $n_{\mathrm{0e}}, n_{\mathrm{1e}}$, and $n_{\mathrm{ae}}$ are the tripod matrix elements,  and
\begin{align}\label{eq:OmegaRMS}
\Omegarms(t_g) &\equiv \sqrt{\frac{1}{t_g}\int_{0}^{t_g}dt\left[|\Omegazeroetilde(t)|^2 + |\Omegaoneetilde(t)|^2 + |\Omegaaetilde(t)|^2\right]}\nonumber\\
&=\Omega_0\sqrt{\frac{1}{t_g}\int_0^{t_g}dt \left[1 + \left(\frac{\ddot{\theta}(t)}{\dot{\theta}^2(t)+\Omega_0^2/4} \right)^2  \right]}.
\end{align}
In the first line of Eq.~\eqref{eq:VRMSSATD}, we have used the fact that 
the terms involving differences of tone frequencies almost exactly average to 0.  Furthermore, in the second line, we have used the fact that our protocol is symmetric  about $t = t_g / 2$, i.e., $\theta(t) = \pi/2 - \theta(t-t_g/2)$ for $t_g/2<t\leq t_g$ [Eq.~\eqref{eq:theta}].

From Eq.~\eqref{eq:VRMSSATD}, we can see that $\tilde{V}_{\rm RMS}$ is related to a more fundamental metric $\Omegarms/2$~[Eq.~\eqref{eq:OmegaRMS}]: the time-averaged RMS value of the instantaneous gap of our Hamiltonian $\Hstirap(t)$ for a SATD-corrected pulse sequence. This metric is solely a property of the ideal tripod Hamiltonian $\Hstirap(t)$ and our SATD pulse sequence. The SATD correction makes the adiabatic gap time dependent, and necessarily increases $\Omegarms(t_g)$ above $\Omega_0$.  

\begin{figure}[t!]
\centering
\includegraphics[width=\linewidth]{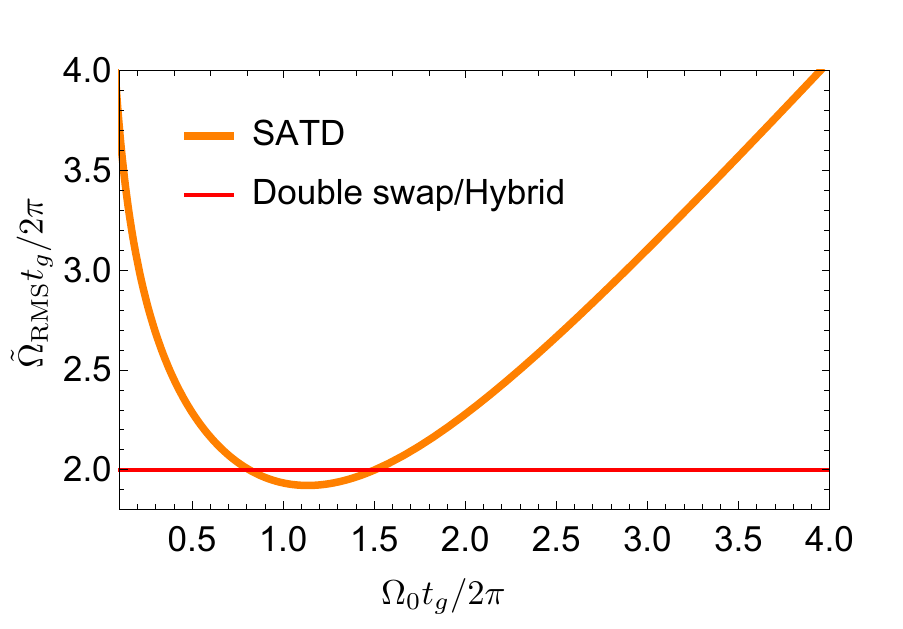}
\caption{$\Omegarms t_g/2\pi$ versus $\Omega_0 t_g/2\pi$ for the SATD (orange curve) as well as the double-swap protocol and the hybrid protocol (red line). For the SATD protocol, $\Omegarms = \Omega_0$ in the adiabatic limit ($\Omega_0 t_g \rightarrow \infty$). Away from the adiabatic limit, the SATD correction increases $\Omegarms$ above $\Omega_0$. The minimum of the SATD plot is $\Omegarms t_g/2\pi = 1.92$, occuring at $\Omega_0t_g/2\pi = 1.135$. The red line $\Omegarms t_g/2\pi =2$  is the value of $\Omegarms$ for the double-swap protocol and the hybrid protocol.}\label{fig:Optimalomegagate}
\end{figure} 

Finally, we can use the property discussed in Sec.~\ref{subsec:InfiniteFamily}:  in the RWA limit, there are an infinite number of SATD protocols that yield a perfect gate fidelity for a given gate time $t_g$.  Out of this set, we choose the protocol that minimizes $\Omegarms$ [and hence approximately minimizes the phase errors arising from the non-RWA energy shifts $\delta \varepsilon_k(t)$ in Eq.~(\ref{eq:energyshift})].  As discussed, the different SATD protocols are indexed by $\Omega_0$, the adiabatic gap associated with the {\it uncorrected} pulses.  We thus seek to identify the value of $\Omega_0 t_g$ that minimizes the SATD energy cost $\Omegarms$.
The behavior of this quantity [obtained numerically for the specific smooth pulse-shape function $P(t)$ given in Eq.~\eqref{eq:Px}] is shown in  Fig.~\ref{fig:Optimalomegagate}.  We find that $\Omegarms / 2\pi$ has a minimum value of approximately $1.92 / t_g$, occurring for $\Omega_0/2\pi = 1.135/t_g$.  We have confirmed that attempting further optimization by using more complex pulse-shape functions $P(t)$ does not yield appreciable improvements.  
We have thus completed step 1 of our two-step approach to enhancing accelerated gates to minimize non-RWA errors:  use the degeneracy of STA protocols to pick a minimum-energy pulse. 

Before proceeding, we pause to note something somewhat remarkable: 
by our optimizing the parameter $\Omega_0$, the SATD pulses are able to achieve an energy-time tradeoff (as quantified by the product 
$\tilde{\Omega}_{\rm RMS} t_g$) that is {\it better} than more obviously fast, nonadiabatic population-transfer protocols (see the red curve in Fig.~\ref{fig:Optimalomegagate}).  These are the so-called double-swap protocols (see e.g., Refs.~\cite{wang2012using,Wang2012using0}) (where one sequentially moves population through the three levels), and the ``hybrid" protocol \cite{Wang2012using0} (a nonadiabatic holonomic protocol).  This point and connections to a formal quantum speed limit valid for time-dependent Hamiltonians \cite{Pires2016Generalized} are discussed in 
Appendix~\ref{sec:lowerbound}.

\subsection{Step 2: Modifying accelerated protocol pulses to cancel non-RWA phase errors}\label{sec:Magnuscorrection}
The next step of our method is to go beyond simply minimizing the non-RWA energy shifts $\delta \varepsilon_k(t)$ in Eq.~(\ref{eq:energyshift}) and actually {\it cancel} them by slightly modifying our SATD pulses.  In keeping with our general philosophy, we derive an analytic prescription for how to do this (as opposed to resorting to a brute-force numerical optimization).

The methodology here is conceptually simple:  to offset the unwanted, time-dependent non-RWA energy shifts, we introduce a time-dependent variation of the central tones in our pulse (i.e.,~a generalized chirp).  Concretely, this means that we introduce a time-dependent shift in each of the three center frequencies $\omega_{j\mathrm{e}}$ ($j = {0,1,\mathrm{a}}$) that appear in our pulse:
\begin{align}\label{eq:modomega}
\omega_{j\text{e}} & \rightarrow \tilde{\omega}_{j\text{e}}(t) = \omega_{j\text{e}} - \delta\varepsilon_{j}(t) +  \delta\varepsilon_{\mathrm{e}}(t),
\end{align}
where $\delta\varepsilon_{j}$ is given by Eq.~(\ref{eq:energyshift}). These frequency shifts ensure that at each instant in time, each tone is resonant (to leading order) with the appropriate transition it is intended to drive. Our chirping procedure does not require any additional calibration procedure:  the matrix elements required to calculate the frequency chirps [see Eq.~\eqref{eq:energyshift}] can be computed using the circuit parameters, parameters that must be accurately estimated even in the absence of chirping.

The net result of our approach is thus a two-step correction to the original pulse in Eq.~(\ref{eq:omegadrive}).  For a given desired gate time $t_g$, we first pick an optimal value of $\Omega_0$ as per Sec.~\ref{subsec:RMSvoltage} and add the SATD correction to the pulses as per Eq.~(\ref{eq:omegaSATD}). Subsequently, we chirp each of the three center frequencies as per 
Eq.~(\ref{eq:modomega}).  We can write the overall modification of each control tone in Eq.~\eqref{eq:V} as  
\begin{align}
V_{j\mathrm{e}}(t) &\,\mathrm{exp}\left(i\omega_{j\mathrm{e}} t\right)
\rightarrow \tilde{V}_{j\mathrm{e}}(t) \,\mathrm{exp}\left(i\int_0^t\tilde{\omega}_{j\mathrm{e}}(t') dt' \right),
\end{align}
where $\tilde{V}_{j\mathrm{e}}(t) = \tilde{\Omega}_{j\mathrm{e}}(t)/n_{j\mathrm{e}}$ are the SATD-corrected pulse envelopes, and $\tilde{\omega}_{j\mathrm{e}}(t)$ are the chirp-corrected central drive frequencies.  

\subsection{Leakage errors and connections to Derivative Removal by Adiabatic Gate (DRAG)}\label{sec:leakage}
Our discussion so far has focused only on errors arising from energy shifts generated by the non-RWA terms in $\Herr(t)$.  There is another generic kind of error to consider:  the non-RWA terms can drive transitions out of the tripod subspace, leading to a final population of nontripod levels.  This kind of error is commonly referred to as leakage, and has been discussed extensively in many other settings (e.g.,~in discussing gate errors in weakly anharmonic transmon-style superconducting qubits \cite{Motzoi2009Simple,Gambetta2011Analytic,Ribeiro2017Systematic}).

The general approach we take in mitigating non-RWA errors partially minimizes leakage by minimizing the size of the control pulses (c.f.~Sec.~\ref{subsec:RMSvoltage}).  However, we do not make any additional modifications of our pulses to further reduce leakage errors.  This is in contrast to our treatment of phase errors (which we further mitigate through frequency chirping).  It is also in contrast to the well-known Derivative Removal by Adiabatic Gate (DRAG) technique \cite{Motzoi2009Simple,Gambetta2011Analytic} for dealing with leakage errors in superconducting circuits driven by a single control tone.  

There are two key rationales for our apparent neglect of pure leakage errors.  The first is purely pragmatic: in general there are many equally important leakage levels, and there is no simple way to modify our pulses (using the Magnus strategy of Ref.~\cite{Ribeiro2017Systematic}) to simultaneously correct all of these error channels. This is because the corrections needed to mitigate leakage are not additive in a simple way:  adding a correction to cancel one leakage transition could make another leakage transition even worse. This is in stark contrast to the usual DRAG problem in superconducting circuits, where there is just a single relevant leakage level (i.e.,~the second excited state of the qubit). The second rationale is that in the perturbative limit (where non-RWA errors are small), leakage errors are much smaller than the phase errors associated with energy shifts.  As shown in Refs.~\cite{wiebe2012improved,roque2020engineering} in the long-gate time limit ($t_g \gg 1/\Deltakl$), phase errors scale as $1/(\Deltakljpm t_g)^2$,  while leakage errors are much weaker, scaling as $1/(\Deltakljpm t_g)^4$.

As we see in the next section (where we implement our ideas in a realistic multilevel superconducting fluxonium circuit), our approach yields extremely good results despite the lack of any specific leakage correction. That said, it would be an interesting topic for future work to devise new methods for mitigating leakage in truly multilevel systems; for example, the general semianalytic method in Ref.~\cite{roque2020engineering} may provide a route for doing this.

\section{Fluxonium qubit}\label{sec:fluxonium}
\subsection{Basic setup}
\begin{figure}[t!]
\centering
\includegraphics[width=0.8\linewidth]{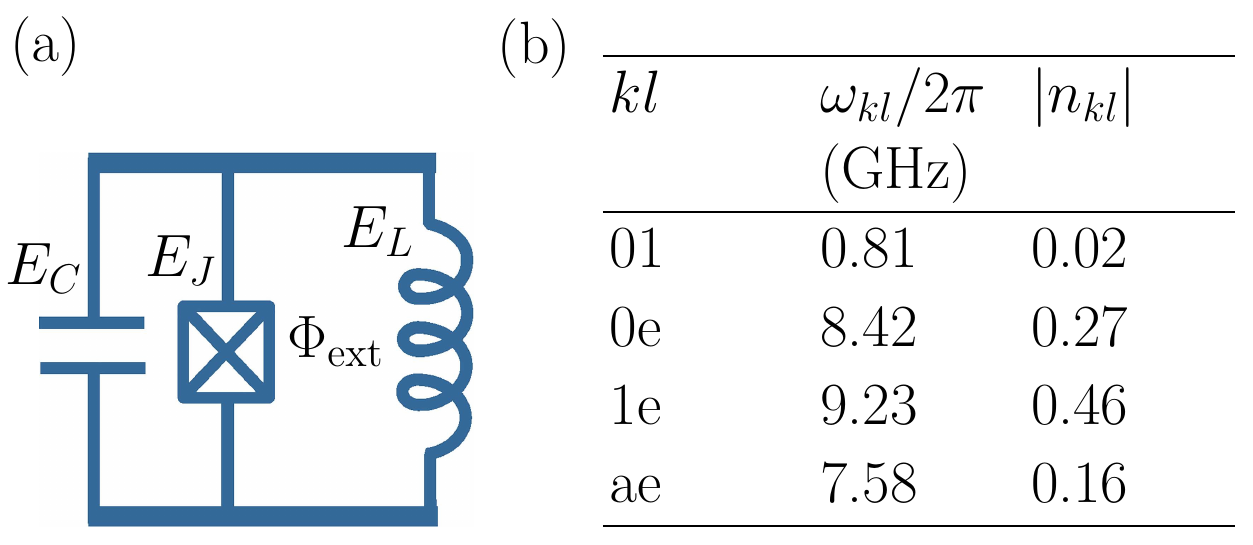}
\caption{(a) Circuit diagram of a fluxonium. (b) The qubit and tripod transition frequencies ($\omega_{kl}$) as well as the magnitude of the charge matrix elements ($n_{kl}$) of the fluxonium whose energy spectrum is given in Fig.~\ref{fig:tripodfluxonium}. The computational states in the tripod system are labeled by $ |0\rangle$, $|1\rangle$, $ |\mathrm{a}\rangle$, and $ |\mathrm{e}\rangle$. The fluxonium parameters used to get the values of $\omega_{kl}$ and $n_{kl}$ in (b) are $E_L/h = 0.063$ GHz, $E_J/h = 9.19$ GHz, $E_C/h = 2$ GHz, and $\Phi_{\mathrm{ext}} = 0.17 \Phi_0$. }\label{fig:schematic}
\end{figure}

The tripod gate we have analyzed is ideally suited to systems where it is difficult to directly drive transitions between qubit levels $\ket{0}$ and $\ket{1}$.  This is often the case in qubits that have long $T_1$ relaxation times.  A paradigmatic example that has received considerable attention recently is a fluxonium-style superconducting qubit~\cite{manucharyan2009fluxonium,Earnest2018Realization,Nguyen2019High,Helin2021Universal}. A fluxonium circuit consists of a single Josephson junction (energy $E_J$) in parallel with both a capacitor
(charging energy $E_C$) and a superinductor  (inductive energy energy $E_L$, typically implemented using a chain of Josephson junctions) [Fig.~\ref{fig:schematic}(a)]. Experiments demonstrate that these qubits can possess exceptionally long relaxation times (on the order of milliseconds)~\cite{Earnest2018Realization,Lin2018Demonstration,Nguyen2019High,Helin2021Universal}. Moreover, they can also be made first-order insensitive to the dephasing from $1/f$ charge noise~\cite{Koch2009Charging}. These properties make a fluxonium an attractive quantum computing platform.  A complication, however, is that the relative isolation of qubit levels (which yields long $T_1$ times) also makes conventional approaches to gates challenging.  This makes our accelerated tripod gate especially attractive.

The fluxonium Hamiltonian is
\begin{equation}\label{eq:Hf}
\hat{H}_f = 4 E_C \hat{n}^2 - E_J \cos(\hat{\varphi} - 2\pi \Phi_{\mathrm{ext}}/\Phi_0) + \frac{1}{2} E_L\hat{\varphi}^2,
\end{equation}
where $\hat{n}$ and $\hat{\varphi}$ are, respectively, the charge and phase operators.  They obey the commutation relation $[\hat{\varphi},\hat{n}] = i$. 
$\Phi_{\mathrm{ext}}$ is the external magnetic flux biasing the loop formed by the Josephson junction and its shunting inductance and $\Phi_0 = h/2e$ is the flux quantum.
Equation~\eqref{eq:Hf} shows that the effective phase potential consists of a cosine potential superimposed on a parabolic background (see Fig.~\ref{fig:tripodfluxonium}).  
The highly tunable, anharmonic nature of the fluxonium allows us to engineer a variety of different candidate four-level tripod systems.   

\subsection{Optimal parameters for a tripod gate}

A first question is to identify parameters yielding an ``optimal" tripod configuration, meaning that we have both a long $T_1$ time and an accelerated SATD gate with small (non-RWA) coherent errors.  
This leads to the following selection criteria:
\begin{enumerate}
\item The ground states should be well isolated from each other (i.e.,~small charge matrix element connecting them) and be nondegenerate.  Strong isolation ensures a long $T_1$ time.
\item The charge matrix elements coupling the excited state to the ground states of the tripod system must be large and have the same order of magnitude. The latter helps minimize coherent errors
arising from non-RWA processes.
\end{enumerate}
Criterion (1) requires us to choose circuit parameters satisfying $E_J \gg E_C$ and $E_{L} \ll E_{J}$ (for well-localized ground states) as well as $0<\Phi_{\mathrm{ext}} \lesssim \Phi_0/4$ (to lift the degeneracy of the ground states). Criterion (2), on the other hand, implies that we have to pick the excited state $|\mathrm{e}\rangle$ of the tripod gate to be the first excited state of the central well.  Furthermore, this state should be delocalized over the potential wells where the ground states $\ket{0}$, $\ket{1}$ and $\ket{\mathrm{a}}$ are located, but should still be somewhat separated from higher-lying energy levels. This implies that the state $\ket{\mathrm{e}}$ must lie near the top edge of the cosine potential, which requires $\sqrt{8 E_{C} E_{J}} \gg 2 E_{J}$. It is obvious that we cannot fulfill criteria 1 and 2 simultaneously in a standard fluxonium circuit.

Since the above requirements cannot be perfectly satisfied simultaneously, we choose parameters that strike an optimal balance,  ensuring that we can end up with both a long-lived qubit and a tripod that allows a high-fidelity SATD gate.  To this end, we perform a numerical search through parameter space to identify optimal regimes~(see Appendix~\ref{sec:justification}). The result is the following near-ideal parameter set for realizing high-fidelity SATD tripod gates in a $T_1$-protected regime: $E_L/h = 0.063$ GHz, $E_J/h = 9.19$ GHz, $E_C/h = 2$ GHz, and $\Phi_{\mathrm{ext}} = 0.17 \Phi_0$ (see Appendix~\ref{sec:justification} for parameter justification). The small inductive energy here puts our device in same regime as the ``Blochnium" circuit recently realized in experiment~\cite{pechenezhskiy2020superconducting}.  

For the above parameter set, the qubit and tripod transition frequencies together with their corresponding charge matrix elements are shown in Fig.~\ref{fig:schematic}(b). The corresponding energy spectrum and wave functions are plotted in Fig.~\ref{fig:tripodfluxonium}. We label the energy levels used for the tripod system by $|0\rangle$, $ |1\rangle$, $ |\mathrm{a}\rangle$, and $|\mathrm{e}\rangle$. The charge matrix element connecting qubit states $|0\rangle$ and $|1\rangle$ is extremely small as desired:  
$|n_{01}| \equiv |\langle 0 | \hat{n} | 1 \rangle| = 0.02$. 
In contrast, the charge matrix elements for the desired tripod transitions are much larger and comparable in magnitude to one another:  $|n_{0\mathrm{e}}|= 0.27$, $|n_{1\mathrm{e}}|= 0.46$ and $|n_{\mathrm{a}\mathrm{e}}|= 0.16$ (right column of the table in Fig.~\ref{fig:schematic}(b)).

\subsection{SATD protocols for tripod gates in fluxonium}
To realize our accelerated tripod gate, we drive the fluxonium circuit with a microwave pulse [described by a voltage $V(t)$] that couples to the charge operator $\hat{n}$.  In the eigenbasis of the bare fluxonium Hamiltonian $\hat{H}_f$ [Eq.~\eqref{eq:Hf}], we can write the Hamiltonian of the driven fluxonium exactly in the general form given in Eq.~\eqref{eq:H}. To realize the accelerated tripod gate, the driving voltage $V(t)$ [Eq.~\eqref{eq:V}] consists of three driving tones $V_{j\mathrm{e}}(t) = \Omega_{j\mathrm{e}}(t)/n_{j\mathrm{e}}$ for $j = 0, 1, \mathrm{a}$ [see Eq.~\eqref{eq:Vme}].
Here $\Omega_{j\mathrm{e}}(t)$ is the complex coupling given in Eq.~\eqref{eq:omegadrive} for the uncorrected pulse and Eq.~\eqref{eq:omegaSATD} for the SATD-corrected pulse. 

\begin{figure}[h!]
\centering
\includegraphics[width=\linewidth]{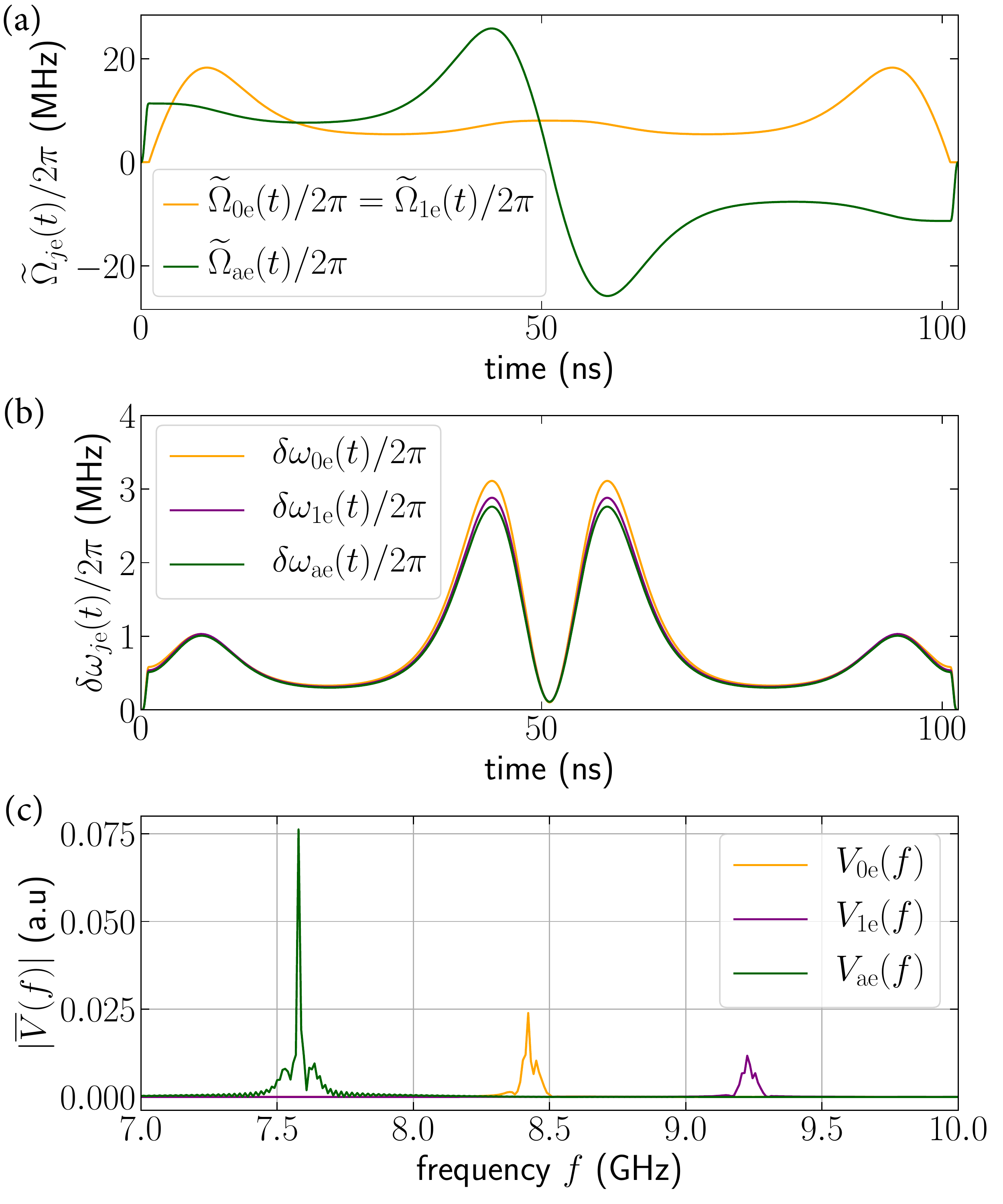}
\caption{ Driving tones used to realize the SATD X gate in a realistic fluxonium. (a) Time profiles of the driving tone envelopes. (b) Time profiles of the frequency chirps $\delta \omega_{j\mathrm{e}}(t) =  \delta\varepsilon_{\mathrm{e}}(t) - \delta\varepsilon_{j}(t)$ of each driving tone. (c) The Fourier transform of the driving tones $|\bar{V}(f)|$ showing distinct peaks corresponding to the tripod transition frequencies shown in Fig.~\ref{fig:schematic}(b). The pulse sequences in (a) are given in Eq.~\eqref{eq:realisticpulse}. They are obtained by our sandwiching the pulses in Eq.~\eqref{eq:omegaSATD} by a ramp time $t_{\mathrm{ramp}}$ at the beginning and end of the protocol, during which the $\tilde{\Omega}_{\mathrm{ae}}(t)$ pulse [green curve in (a)] is turned on and off, respectively, using the smooth polynomial function given in Eq.~\eqref{eq:Px}. The parameters used are $t_g=100$ ns, turn-on and turn-off time $t_{\mathrm{ramp}} = 1$ ns, and $\Omega_0/2\pi = 1.135/t_g = 11.35$ MHz.  The parameter set for the fluxonium is the same as that used for Fig.~\ref{fig:schematic}.}
\label{fig:pulse}
\end{figure} 
The envelopes of the driving tones $\tilde{\Omega}_{j\mathrm{e}}(t)$ of the SATD tripod gate pulse (for optimal $\Omega_0 t_g$) are shown in Fig~\ref{fig:pulse}(a).  We have slightly modified the pulses derived in Sec.~\ref{sec:accelerated} so that $V(t)$ goes smoothly to zero at the start and end of the protocol (as would be in the case in experiment).  We do this by sandwiching the ideal pulses with a short ramp up (down) of duration $t_{\mathrm{ramp}} = 0.01 t_g$ at the beginning (end) of the protocol.  During this ramp, the $\tilde{\Omega}_{\mathrm{ae}}(t)$ tone is smoothly turned on and off, respectively (see Appendix~\ref{sec:ramp}). We specifically use Eq.~\eqref{eq:Px} for a smooth ramp function that turns the pulse on and off. Including these ramps does not appreciably change our results.

Figure~\ref{fig:pulse}(b) shows the size of the frequency chirps $\delta\omega_{j\mathrm{e}}(t$) applied to each driving tone [$\omega_{je} \rightarrow \omega_{je} + \delta\omega_{j\mathrm{e}}(t)$ for $j = 0,1,\mathrm{a}$], to correct for the energy shift of the computational levels due to the non-RWA terms. The sizes of these corrections $ \delta\omega_{j\mathrm{e}}(t)$ are on the order of MHz, while the base frequency $\omega_{je}$ is on the order of GHz. The Fourier transform of the driving pulse is shown in Fig.~\ref{fig:pulse}(c), where each distinct peak in the plot corresponds to one of the tripod transition frequencies whose values are given in the table in Fig.~\ref{fig:schematic}(b).   

\section{Gate performance: comparing different error channels}\label{sec:gateperformance}
To quantify the performance of our accelerated tripod gate, we calculate the state-averaged fidelity of the gate.  This is given by~\cite{bowdrey2002fidelity}
\begin{equation}\label{eq:fidelity}
\bar{F} = \frac{1}{6}\sum_{m = \pm x, \pm y,\pm z}\mathrm{Tr}\left[\hat{U}_{\mathrm{q}} \hat{\rho}_m\hat{U}_{\mathrm{q}}^\dagger\hat{\rho}_m(t_g)   \right],
\end{equation}
where $\hat{\rho}_m $ is an axial pure state on the qubit's Bloch sphere with $m \in \{\pm x, \pm y, \pm z \}$ [e.g.~$\hat{\rho}_x = 1/2(|0\rangle + |1\rangle)(\langle 0| + \langle 1|)$].   $\hat{\rho}_m(t_g) $ is the laboratory-frame density matrix of the system at the end of the protocol ($t = t_g$) for the initial state $\hat{\rho}_m$.  Here $\hat{U}_{\mathrm{q}}$  
is a product of the ideal target unitary gate operation in the qubit subspace $\hat{U}_{\rm G,01}$
[see~Eq.~\eqref{eq:UQubitAdiabatic}]
and an innocuous phase factor corresponding to dynamical phases in the laboratory frame:
\begin{equation}
\hat{U}_{\mathrm{q}} =  \hat{U}_\mathrm{diag,01}(t_g)  \hat{U}_\mathrm{G,01},
\end{equation}
where $\hat{U}_\mathrm{diag,01}(t_g) = \sum_{k=0,1}e^{-i\int_{0}^{t_g}\varepsilon_k(t) dt} |k\rangle \langle k|$.  Here  $\varepsilon_k(t) = \varepsilon_k$  for unchirped protocols and $\varepsilon_k(t) = \varepsilon_k + \delta\varepsilon_k(t)$ for chirped protocols. In what follows, we use this standard metric to characterize a target X qubit gate in the presence of both coherent (non-RWA) errors and dissipation.  

\subsection{Effects of coherent errors only}
Consider first the case where dissipation is ignored, and the only sources of gate errors are the non-RWA terms in Eq.~\eqref{eq:HerrIntro}.  To calculate gate performance in this limit, we numerically evolve initial states as per the laboratory-frame Hamiltonian~[Eq.~\eqref{eq:H}] using the Python package QuTiP~\cite{johansson2012qutip,johansson2012qutip2}. We perform the simulation by including the 18 lowest energy levels of the fluxonium
and all charge matrix elements in this space.  We check that including more energy levels in the simulations does not change the results.

\begin{figure}[h!]
\centering
\includegraphics[width=\linewidth]{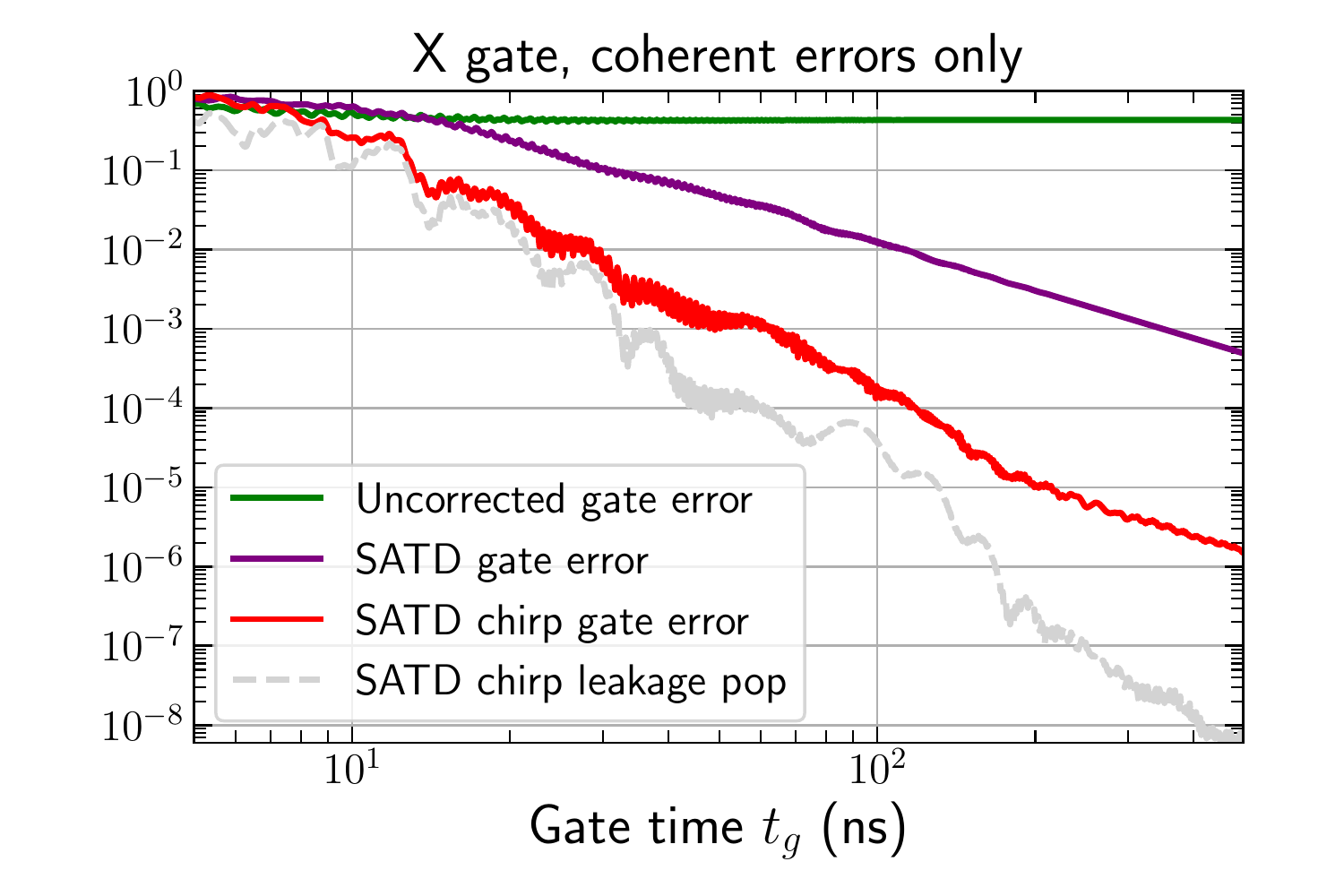}
\caption{State-averaged gate error $\bar{\varepsilon} = 1-\bar{F}$ (solid curves) versus gate time $t_g$ for different realizations of a tripod X gate in a fluxonium qubit in the absence of dissipation.  We keep the lowest 18 levels of the circuit and include all non-RWA error channels. The curves correspond to uncorrected adiabatic pulses (green), SATD pulses without frequency chirping (purple) and SATD pulses with frequency chirping (red).  For all pulses, we use the optimal value of the uncorrected gap frequency $\Omega_0/2\pi = 1.135 /t_g$  that minimizes the RMS voltage. Also shown is the leakage population outside the tripod subspace for the chirped SATD protocol (dashed gray curve).
}\label{fig:Errormapchirp}
\end{figure}

In Fig.~\ref{fig:Errormapchirp}, we plot the gate errors $\bar{\varepsilon} = 1- \bar{F}$ for a target tripod gate $\hat{U}_{\mathrm{G},01} = -\hat{\sigma}_{x,01}$ (an X gate) obtained for different choices of pulses.  For all curves, the uncorrected gap frequency at each protocol time $t_g$ is picked to be the optimal value $\Omega_0/2\pi = 1.135/t_g$ (see Sec.~\ref{subsec:RMSvoltage}).  If we use the uncorrected adiabatic pulses  [see~Eq.~\eqref{eq:omegadrive}; green curve], errors arise both from nonadiabatic transitions and from non-RWA processes.  If instead we use the SATD-corrected pulses [see~Eq.~\eqref{eq:omegaSATD}; purple curve], nonadiabatic errors are completely canceled, leaving only non-RWA errors.  Finally, if we also frequency chirp the SATD pulses as per Eq.~\eqref{eq:modomega} (red curve), we further reduce gate errors by reducing the leading non-RWA errors.  This yields a dramatic improvement over the unchirped SATD protocol at long gate times.  

As discussed, our corrections do not specifically cancel pure leakage errors.  To characterize these, we calculate the state-averaged population outside the tripod subspace (leakage population) at the end of the protocol.  This is given by
\begin{equation}
    1 - \bar{\mathcal{P}}_{\mathrm{tripod}} = 1 - \frac{1}{6}\sum_{m = \pm x, \pm y,\pm z}\mathrm{Tr}\left[ \hat{P}_{\mathrm{tripod}}\hat{\rho}_m(t_g)   \right],
\end{equation}
where $\bar{\mathcal{P}}_{\mathrm{tripod}}$ and $\hat{P}_{\mathrm{tripod}}$ are, respectively, the state-averaged population and the projector in the tripod subspace. The final state-averaged leakage population is plotted as a dashed gray curve in  Fig.~\ref{fig:Errormapchirp}.  As discussed, leakage makes a minimal contribution to the error at moderate to long gate times.  

\subsection{Effects of $1/f$ dephasing noise only}
\subsubsection{Effective modeling of non-Markovian noise}\label{sec:fluxnoisemodel}
We now turn to modeling additional gate errors arising from dissipation.This will give rise to a nontrivial competition:  mitigating coherent non-RWA errors favors long gate times (and hence low powers), whereas minimizing dissipative errors favors short gate times. Given our operating point (isolated qubit states, but not at the flux sweet spot), $1/f$ flux noise  will often be the dominant dissipation mechanism.
In what follows, we thus focus on dephasing dissipation and the resulting competition between coherent and dissipative errors.  In the relevant basis of dressed states, dephasing noise can cause transitions; hence, for our scheme, there is no strong qualitative difference between the effects of dephasing and $T_1$ relaxation.   
That said, the additional effect of explicit $T_1$ relaxation is analyzed in detail in Appendix~\ref{sec:ErrorT1}.

As is common 
\cite{Makhlin2004Dephasing,Ithier2005Decoherence,groszkowski2018coherence,Didier2019,di2019efficient}, we model the $1/f$ non-Markovian noise using an approximate Markovian description that qualitatively captures the relevant dephasing timescales correctly.
Letting $\hat{\rho}(t)$ denote the fluxonium reduced density matrix, we model our system by the Lindblad-form master equation
\begin{equation}\label{eq:Lindblad}
    \partial_t\hat{\rho}(t) = -i [\hat{H}(t), \hat{\rho}(t)] + \left(
        \hat{Z}\hat{\rho}\hat{Z} - \frac{1}{2}\{ \hat{Z}^2,\hat{\rho}\} \right).
\end{equation}
$\hat{H}(t)$ is the driven fluxonium Hamiltonian [Eq.~\eqref{eq:H}],
and the Hermitian operator $\hat{Z}$ has the general form
\begin{equation}\label{eq:c0}
    \hat{Z} = \sum_k\mathrm{sgn}\left(\frac{\partial \varepsilon_{k}}{\partial \Phi_{\mathrm{ext}}} \right)  \sqrt{2\Gamma_k}|k\rangle \langle k|.
\end{equation}
Heuristically, this describes the fact that each fluxonium energy level $\varepsilon_{k}$ depends on the bias flux, and hence flux noise causes each energy to fluctuate.  We have written the coupling constant associated with each level $\ket{k}$ in terms of an overall sign (which captures whether $\varepsilon_k$ increases or decreases with increasing flux) and a magnitude $\Gamma_k$.

To fix the couplings $\Gamma_k$, we use the fact that for classical, Gaussian $1/f$ noise, free induction decay of a given coherence 
$\rho_{kl} \equiv \bra{k} \hat{\rho} \ket{l} $ has a decay envelope of the form $\exp \left[-(t/T_{\varphi,kl})^2\right]$ (up to logarithmic corrections).  A standard calculation (see e.g., Ref.~\cite{groszkowski2018coherence}) yields:
\begin{equation}
    \label{eq:dephasingrate}
    1/T_{\varphi,kl} = A_{\Phi_{\mathrm{ext}}}
    |\partial_{\Phi_{\mathrm{ext}}}
    (\varepsilon_k - \varepsilon_l)|
    \sqrt{|\mathrm{ln} \, D|}.
\end{equation}
Properties of the flux-noise spectral density enter only through $A_{\Phi_{\mathrm{ext}}}$ (the standardly defined flux-noise amplitude) and $D$ (the product of the measurement time and the low-frequency cutoff of the noise; see Refs.~\cite{Mohamed2020Universal,groszkowski2018coherence}).

We construct our approximate Markovian master equation by picking the couplings $\Gamma_k$ to ensure that for free induction decay over a time $t_g$, the {\it final} decay of a large set of coherences is captured correctly.  In particular, we take $\ket{1}$ as a reference level (i.e.,~the lowest energy level of the fluxonium), and insist that 
the net decay of any coherence $\rho_{k1}$ ($k \neq 1$) after an evolution time $t_g$ is the {\it same} for our Markovian dynamics as it would be following the non-Markovian, Gaussian-lineshape decay described by Eq.~(\ref{eq:dephasingrate}).  This leads to the choice $\Gamma_{1}=0$, and for $k \neq 1$: 
\begin{equation}
    \label{eq:gammak}
    \Gamma_k = \Gamma_k(t_g) =  t_g/ (T_{\varphi, {k1}})^2.
\end{equation}
We stress that $\Gamma_k$ (and hence our master equation) depends only on the choice of total evolution time $t_g$; during the evolution the dephasing superoperator is constant.  

The above choice guarantees that at the end of evolution for a time $t_g$, the overall free induction decay of any coherence involving the qubit level $\ket{1}$ (the lowest energy state of the circuit) is captured correctly.  As discussed in Appendix~\ref{sec:inconsistency}, there is no way to make a choice for $\Gamma_k$ that captures the decay of all coherences correctly.  Nonetheless, as shown in Appendix~\ref{sec:inconsistency}, our approach if anything overestimates the dominant dephasing within the tripod subspace
(see Table~\ref{table:dephasing} in Appendix~\ref{sec:inconsistency}).  Note also that our approach overestimates dephasing compared with alternative approximations that use an explicitly time-dependent $\Gamma_k$~\cite{Didier2019,di2019efficient}.  Moreover, our modeling of dephasing as being Markovian is also a worst-case scenario, as there is spectral weight for arbitrary transitions (which would not be true for realistic $1/f$ noise where there is little spectral weight to drive transitions between levels with large energy detunings).

In what follows we take $A_{\Phi_{\mathrm{ext}}} = 3\mu\Phi_0$ (which is typical for state-of-the-art experiments~\cite{yan2016flux,Nguyen2019High,Helin2021Universal}), and also choose 
(following Refs.~\cite{Koch2007Charge,Mohamed2020Universal,groszkowski2018coherence}) $D = 
(2 \pi \times 1 \,{\rm Hz}) \times 10\, \mu s$.  With these choices and for the circuit parameters used here, we find dephasing times  $T_{\varphi,kl}$ that are on the order of approximately $\sim 1- 100 \mu$s~[Table~\ref{table:dephasing} in Appendix~\ref{sec:inconsistency}].  The dephasing of the qubit levels is $T_{\varphi,01} = 7$ $\mu$s. Our results are largely unchanged if one adds $T_1$ decay processes (e.g., dielectric loss with dielectric quality factor  $Q_{\mathrm{diel}} \gtrsim 5 \times 10^6$); see Appendix~\ref{sec:ErrorT1}. 

\subsubsection{RWA gate performance in the presence of dephasing noise}

\begin{figure}[h!]
\centering
\includegraphics[width=\linewidth]{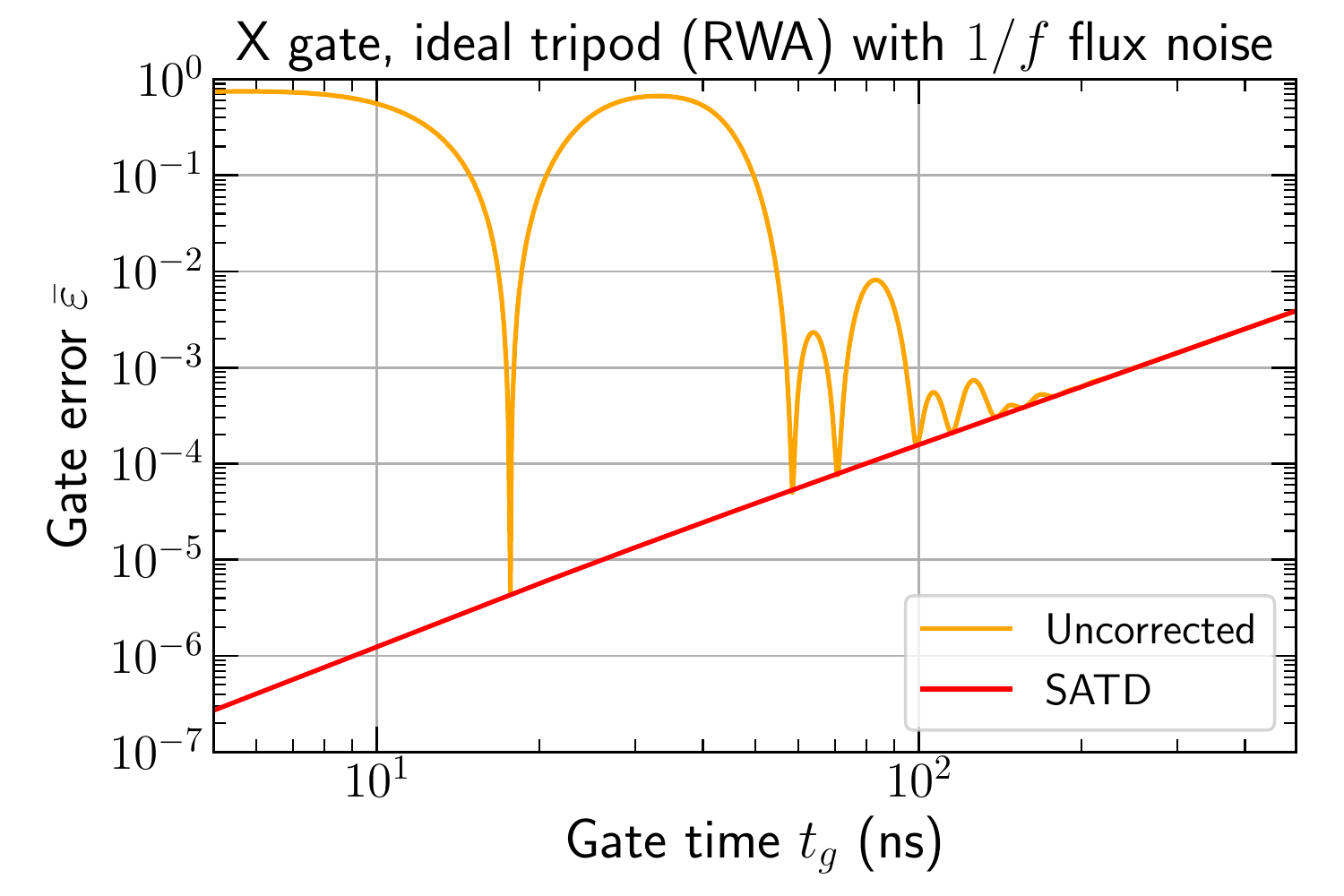}
\caption{State-averaged gate error $\bar{\varepsilon} = 1-\bar{F}$ as a function of gate time $t_g$ for ideal (four-level, RWA) tripod X gates.  We compare uncorrected adiabatic pulses (orange curve) against the accelerated SATD pulses (red curve) in the presence of $1/f$ flux noise.  
The scale of the uncorrected pulses is set by $\Omega_0/2\pi = 100$ MHz, and the $1/f$ flux noise has a strength parameterized by $A_{\Phi_{\mathrm{ext}}} = 3 \mu \Phi_0$ (yielding a qubit dephasing time $T_{\varphi,01} = 7$ $\mu$s).  Fluxonium parameters match those given in the caption for Fig.~\ref{fig:schematic}.}
\label{fig:Errormap_ideal}
\end{figure}

We numerically evolve the master equation Eq.~\eqref{eq:Lindblad} 
(with the above form for the dephasing superoperator) and use the results to calculate the state-averaged gate error $\bar{\varepsilon} = 1- \bar{F}$ of a tripod X gate [see~Eq.~\eqref{eq:fidelity}].  We first consider the case where all non-RWA terms in the coherent Hamiltonian are ignored, so that errors are the result of only dissipation or nonadiabaticity.  The results as a function of gate time $t_g$ are shown in  Fig.~\ref{fig:Errormap_ideal}, where the performances of the uncorrected and accelerated (SATD) pulses are compared. The simulations of the gate dynamics for all gate times $t_g$ are done for  fixed $\Omega_0/2\pi = 100$ MHz.

The behavior here is generic and as expected.  
In the adiabatic regime where $t_g \gg 1/\Omega_0$, the uncorrected and SATD gates have almost identical performance.  In this regime the error $\bar{\varepsilon}$ is dominated by dephasing, and grows quadratically with $t_g$ (reflecting the quadratic loss of coherence expected from $1/f$ noise at short times). 
This quadratic-in-$t_g$ error scaling continues down to small values of $t_g$ for the SATD curve, as in this case, dephasing is the only error mechanism.  In contrast, the uncorrected curve has much larger errors at short time, corresponding to nonadiabatic errors.  There are special sharply defined values of $t_g$  where these nonadiabatic errors constructively cancel; as discussed in Ref.~\cite{Ribeiro2019Accelerated}, these are difficult to exploit experimentally as they require extreme fine-tuning.  

\section{Full gate performance and comparison
against brute-force direct driving}
\label{sec:directdriving}

\subsection{Direct driving gate and power scaling}\label{sec:powerscale}

Having investigated the impact of different error channels (nonadiabatic errors, non-RWA errors, and dephasing), we are now ready to study the accelerated tripod gate with all error channels present.  To properly understand the advantages of our  accelerated tripod gate, it is instructive to compare its performance against that of gates realized using other approaches. For example, one can consider comparing it against a traditional Raman gate (realized by indirectly driving the qubit states via off-resonant coupling to a common excited state ~\cite{Vitanov2017Stimulated}) or against ``direct driving'' (DD) gates (where one resonantly drives the qubit transition between $\ket{0}$ and $\ket{1}$). One key
advantage of the SATD tripod approach over these alternatives is their intrinsic robustness:  the SATD gate inherits the resilience to control-pulse imperfections that would be expected by a purely adiabatic gate.  For example, there is a marked resilience to errors in the magnitude of $\Omega_0$, which could arise either from imperfect pulse calibration or from uncertainties in matrix elements.  This advantage persists even in the accelerated adiabatic regime~\cite{Ribeiro2019Accelerated}.

In what follows we demonstrate {\it additional} advantages of our tripod gate, by specifically comparing its performance against that of the simple DD gate (see Appendix~\ref{sec:Raman} for  comparisons with Raman gates). To this end, we note that as we are working with a $T_1$-protected qubit, the magnitude of the matrix element used to drive the qubit transition in the DD gate ($|n_{10}|$) is small; however, for a large enough power one could, in principle, still achieve a given gate.  
To compare our tripod approach against the DD gate, we want to compare not only the gate error $\bar{\varepsilon}$ but also the {\it power} required to achieve the gate.  As we see in what follows, in many experimental systems, additional constraints limit the magnitude of pulses that can be used.  This will provide a strong advantage in many regimes for the tripod gate.  

We begin by writing
 a simple pulse shape that can be used to realize a DD gate~\cite{roque2020engineering}: 
\begin{equation}\label{eq:Vdirect}
 \tilde{V}_{\mathrm{DD}}(t) =  \frac{\chi}{t_g |n_{01}|}\left[1- \cos\left(\frac{2\pi t}{t_g} \right) \right]\cos\left[\int_0^{t}dt'\,\tilde{\omega}_{01}(t')\right],
\end{equation}
 where $\chi = \pi$ for X gate.  Similarly to our accelerated adiabatic gate, we perform chirping to partially correct for non-RWA errors, and hence 
$\tilde{\omega}_{01}(t) = \omega_{01} + \delta \varepsilon_{1}(t) - \delta \varepsilon_{0}(t)$ [see~Eq.~\eqref{eq:modomega}]. 

Using the definition in Eq.~(\ref{eq:VRMS}), we can calculate the RMS time-averaged voltage of the DD gate.  We can also use our previous result [Eq.~\eqref{eq:VRMSSATD}] for the RMS time-averaged voltage of the SATD gate (using the optimal value of $\Omega_0 t_g$ discussed in Sec.~\ref{subsec:RMSvoltage}).  These two RMS voltages are given by
\begin{subequations}
    \begin{align}
    \tilde{V}_{\mathrm{RMS,DD}} &= \frac{\sqrt{3}\chi}{2|n_{01}|t_{g,\mathrm{DD}}},
    \label{eq:VRMSDD} \\
    \tilde{V}_{\mathrm{RMS,SATD}} &=\frac{1.92 \pi }{t_{g,\mathrm{SATD}}} \sqrt{\frac{\cos^2\alpha}{|n_{\mathrm{0e}}|^2} + \frac{\sin^2\alpha}{|n_{\mathrm{1e}}|^2} + \frac{1}{|n_{\mathrm{ae}}|^2}}\label{eq:VRMSSATD1}.
\end{align}
\end{subequations}
In both cases, the RMS voltage scales inversely with the gate time $t_g$, but note the crucial dependence on matrix elements and gate type.  

It follows from Eq.~(\ref{eq:VRMSDD}) that an X gate is the most energy-consuming (and hence problematic) gate if one uses the DD approach.  We thus focus on this gate in what follows, and compare it against our accelerated adiabatic approach.
Using the matrix elements for the fluxonium parameters used throughout this paper [see Fig.~\ref{fig:schematic}(b)], we  get $\VRMS$ for the X gate ($\chi =\pi$ for the DD approach and $\alpha = \pi/4$ for the tripod approach) as
\begin{subequations}\label{eq:VRMS}
\begin{align}
    \tilde{V}_{\mathrm{RMS,DD}}&= \frac{136}{t_{g,\mathrm{direct}}},\label{eq:Vdirect1}\\
    \tilde{V}_{\mathrm{RMS,SATD}} &= \frac{42.1}{t_{g,\mathrm{SATD}}}.\label{eq:VSATD1}
\end{align}
\end{subequations}

The equations already describe a crucial advantage of the tripod approach over direct driving:  for a fixed gate time, the SATD tripod approach requires drive amplitudes that are 3.2 times smaller, corresponding to a factor of $9$ or greater saving in power.  An immediate consequence is that if we fix the RMS voltage to be the same for both protocols, the SATD tripod gate will be faster.

We stress that a constraint on the size of drive amplitudes emerges naturally in typical superconducting circuits, where drives are applied via a sufficiently detuned microwave cavity mode (e.g., $\omegar/2\pi = 2$ GHz; see Appendix~\ref{sec:Tonecav} for a justification) that couples to the circuit as
\begin{equation}\label{eq:Hcoupling}
\hat{H}_{\mathrm{c}} = \sum_{kl} g n_{kl}|k\rangle \langle l|(\hata^\dagger + \hata).
\end{equation}
Here $g$ is the cavity-fluxonium coupling strength, $n_{kl}$ is a charge matrix element, and $\hat{a}$ is the cavity photon annihilation operator.  A standard constraint in such setups is that the time-averaged intracavity photon number $\bar{n}_{\rm cav}$ should not exceed some small value to avoid additional dissipative mechanisms; here we require $\bar{n}_{\rm cav} \leq 0.05$.  This in turn directly constrains the RMS voltage via the relation (see Appendix~\ref{sec:resonatorfull})
\begin{equation}\label{eq:VRMSg}
 \tilde{V}_{\mathrm{RMS}}^2 = 2  \bar{n}_{\mathrm{cav} } g^2
 \leq 0.1 g^2.
\end{equation}

Instead of constraining the cavity photon number, one could consider minimizing some other quantities that are sensitive to the power delivered to the system.  This is expected to yield results qualitatively similar to those obtained with our chosen constraint on  $\bar{n}_{\mathrm{cav}}$, which is motivated by experimental observations as well as other theoretical studies (e.g., Ref.~\cite{Mohamed2020Universal}).
We stress that the motivation for limiting power is empirical, and is not based on some rigorous theory of large-drive dissipation.  A variety of experiments exhibit performance degradation when high-amplitude drive pulses are used (see, e.g., Refs.~\cite{Earnest2018Realization,Sank2016Measurement,minev2019catch,Lescanne2019Escape}). Many mechanisms could contribute to these observations, including a simple heating of the circuit environment induced by high drive power.  The specific nature of the mechanism is not relevant to the results we present below.    

Finally, we remark that since the qubit frequency is lower than  the drive frequencies of our tripod gate, instead of driving the qubit via a cavity, one could consider coupling the fluxonium circuit directly to a transmission line that includes a high-pass filter~\cite{Referee}. Such a scenario would not only be less constraining on the amplitudes of the high-frequency tones required to perform our tripod gate but would also still protect the qubit subspace from the undesired transitions caused by the low(er)-frequency fluctuations of the transmission line. Nevertheless, here, we concentrate on studying the case of driving the qubit indirectly via a cavity, as that is the most widely used approach in experiments, and hence might be the easiest thing to try in a first experimental realization of our gate.

In what follows, we compare the SATD tripod gate against the DD gate for fixed values of $g$ (and hence fixed maximum possible $\VRMS$).  We see that this physically motivated power constraint gives the accelerated adiabatic gate an important advantage. Further details about driving via a cavity (including the driving pulse applied to the cavity and constraints that allow one to ignore cavity-induced dissipation) are given in   
Appendix~\ref{sec:resonatorfull}. 

\subsection{Comparison of gate performance, coherent errors only}
\begin{figure}[t!]
\centering
\includegraphics[width=\linewidth]{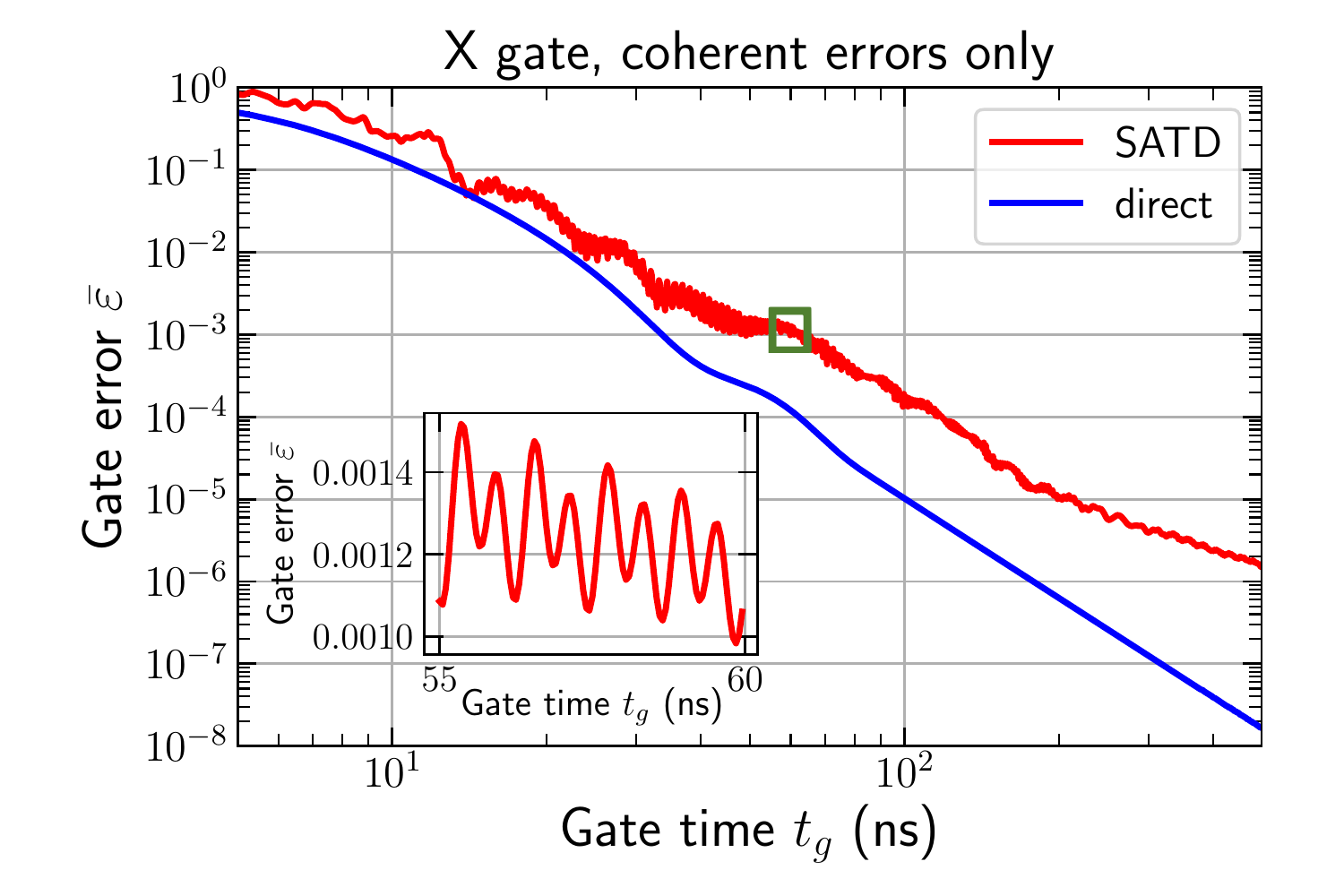}
\caption{State-averaged gate error $\bar{\varepsilon} = 1-\bar{F}$ as a function of gate time $t_g$, for SATD tripod (red curve) and direct-driving (blue curve) realizations of a fluxonium X gate.  Dissipation is not included here, but non-RWA error channels are.  The inset shows an enlargement of the SATD gate error. The gates are calculated with the 18 lowest energy levels of the circuit. For the SATD gate, we use the optimal uncorrected gap frequency $\Omega_0/2\pi = 1.135/t_g$ that minimizes the RMS voltage at each gate time $t_g$. 
The fluxonium parameters are as for Fig.~\ref{fig:schematic}.}\label{fig:Error_nodephasecrop}
\end{figure}

We first compare the accelerated tripod gate to the DD gate in the absence of dissipation, but including all non-RWA terms.  
Figure~\ref{fig:Error_nodephasecrop} shows the 
state-averaged gate errors $\bar{\varepsilon}$ for an X gate as a function of gate time $t_g$ for our chosen fluxonium parameters.  To mitigate non-RWA errors, the SATD protocol is implemented with use of the optimal value of $\Omega_0$ (see Sec.~\ref{subsec:RMSvoltage}) and is frequency chirped (see Sec.~\ref{sec:Magnuscorrection}); the DD gate is also chirped.

\begin{figure*}[t!]
\centering
\includegraphics[width=\linewidth]{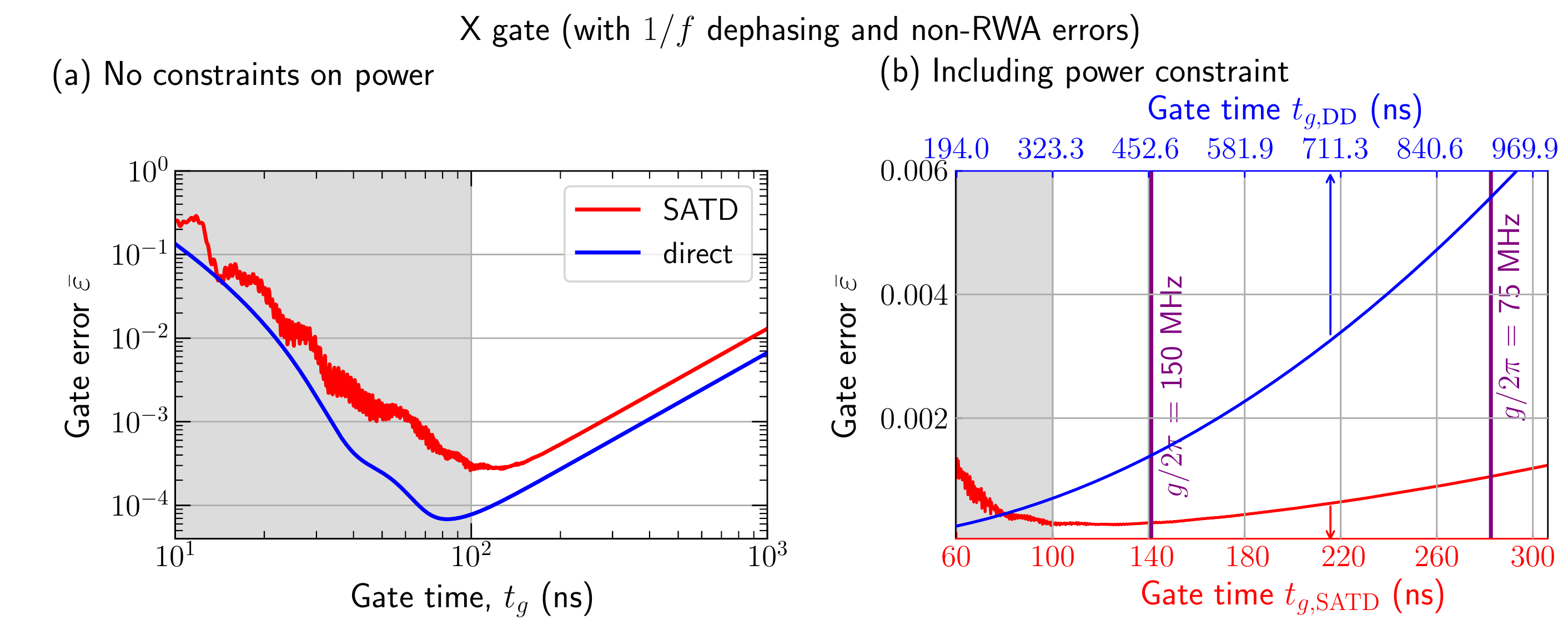}
\caption{
Comparison between state-averaged X-gate error $\bar{\varepsilon} = 1-\bar{F}$ for SATD (red curve) and direct-driving (blue curve) protocols
in our fluxonium system, including effects of $1/f$ flux noise dephasing as well as non-RWA errors; 18 fluxonium levels are included in the simulations.  Dephasing is treated as per Sec.~\ref{sec:fluxnoisemodel} with a $1/f$ flux noise amplitude $A_{\Phi_{\mathrm{ext}}} = 3 \mu \Phi_0$ corresponding to the qubit dephasing time  $T_{\varphi,01} = 7$ $\mu$s.
(a)  Log-log plot of gate error $\bar{\varepsilon}$ versus gate time $t_g$.  For each time, the DD protocol has a smaller error but uses significantly larger pulse amplitudes.  (b)  Linear plot of gate error $\bar{\varepsilon}$ for both protocols at equal power levels $\VRMS$ [see~Eq.~(\ref{eq:VRMS})].  The shaded (unshaded) region corresponds to the time regime where the gate errors are dominated by non-RWA errors (dephasing). For the same value of $\VRMS$, the gate time for the DD protocol is longer than for the SATD protocol (hence the two distinct $x$ axes).  Vertical lines indicate the shortest gate times possible (corresponding to the maximum allowed $\VRMS$; see Eq.~\eqref{eq:VRMSg}) for cavity-based driving for time-averaged cavity photon number $\bar{n}_{\mathrm{cav}} = 0.05$ and a fixed cavity-fluxonium coupling $g$ (purple vertical lines).  Fluxonium parameters are the same as for Fig.~\ref{fig:schematic}.   
}\label{fig:Error_dephasecrop}
\end{figure*} 

We see that for both approaches, errors increase as $t_g$ is reduced.  This simply reflects the higher drive amplitudes needed at shorter times (which in turn increase non-RWA errors).  For all gate times, the coherent errors are larger for SATD versus DD.  This is a result of the SATD protocol using multiple drive tones, and being subject to more near-resonant non-RWA error channels (including those involving higher-energy excited states).  Interference between its multiple drive tones causes the SATD error curve to exhibit fast, low-amplitude oscillations as a function of $t_g$ (see the inset in Fig.~\ref{fig:Error_nodephasecrop}).

The results in Fig.~\ref{fig:Error_nodephasecrop} may seem depressing.  However, as we argue in what follows, they are misleading.  For a given gate time, we have already seen (see~Sec.~\ref{sec:directdriving}) that DD requires considerably larger pulse amplitudes than SATD.  Once we enforce power constraints associated with realistic driving via a cavity (and also include dissipation), the accelerated tripod gate will have a marked advantage.

\subsection{Qubit gates with $1/f$ flux noise}\label{sec:fnoise}

We next compare the DD gate and accelerated SATD tripod gate including $1/f$ flux noise (modeled as per Sec.~\ref{sec:fluxnoisemodel}) as well as non-RWA errors; the results are shown in Fig.~\ref{fig:Error_dephasecrop}(a).  For both protocols, the error is nonmonotonic with gate time $t_g$.  For short times (gray shaded region), the errors are dominated by non-RWA effects and decrease with increasing $t_g$.  For longer times dephasing dominates, causing the error to increase with increasing $t_g$.  In this latter regime, both curves increase quadratically, but the SATD curve has the higher error.  This corresponds to a shorter decoherence timescale for tripod coherences (as used in SATD) versus the qubit $01$ coherence (see Table~\ref{table:dephasing} in Appendix~\ref{sec:inconsistency}).

In Fig.~\ref{fig:Error_dephasecrop}(b) we replot these results in a way that now accounts for power constraints that arise when driving is performed through a cavity (see~Sec.\ref{sec:directdriving}).  Each vertical cut corresponds to a fixed value of the RMS voltage (see.~Eq.~\eqref{eq:VRMS}); given that DD uses more power, fixing the voltage thus results in $t_g$ for SATD approximately $3.2$ smaller for SATD (for our parameter choices) than for DD (hence the different $t_g$ axes for DD versus SATD).  We thus see that once power is constrained (through $\VRMS$), the accelerated tripod gate has a marked advantage over DD for $t_g \gtrsim 100$ ns; this is the regime where dephasing dominates non-RWA errors, and hence the faster speed of SATD is advantageous. 

We can quantify this relative advantage by the 
ratio of DD and SATD errors for the same fixed $\VRMS$:
\begin{equation}\label{eq:zeta}
\zeta \equiv \frac{\bar{\varepsilon}_{\mathrm{DD}}({\tilde{V}_{\mathrm{RMS}})}} {\bar{\varepsilon}_{\mathrm{SATD}}(\tilde{V}_{\mathrm{RMS}})}.
\end{equation}
In the dephasing-limited regime, the gate errors for the SATD and DD gates both exhibit a quadratic scaling with $t_g$.  For our chosen parameters and $1/f$ flux noise strength, we find
\begin{equation}
    \zeta = \frac{\bar{\varepsilon}_{\mathrm{DD}}}{ \bar{\varepsilon}_{\mathrm{SATD}}} = 5.3. 
\end{equation}

We thus have a central conclusion of our work:  our analytically designed accelerated tripod gate allows one to suppress gate errors 
(associated with an X gate) by more than a factor of $5$ compared with the more simplistic DD approach. 
This conclusion holds for $\VRMS$ values small enough that errors are dominated by dephasing.    
As discussed in Sec.\ref{sec:directdriving}, the maximum value of $\VRMS$ is determined by the cavity-qubit coupling $g$ and the requirement that the time-averaged intracavity photon number $\bar{n}_{\rm cav} \leq 0.05$.  As shown in Fig.~\ref{fig:Error_dephasecrop} (purple vertical lines), realistic choices of $g$ put us squarely in this dephasing-dominated regime where we have a strong advantage.  

While we have focused on comparing our accelerated tripod gate against DD, it is also worthwhile to consider its absolute performance.  As shown in Fig.~\ref{fig:Error_dephasecrop}(a), we are able to achieve a fidelity of approximately $0.9997$ in gate time $t_g = 100$ ns (including $1/f$ flux noise with $A_{\Phi_{\mathrm{ext}}} = 3\mu\Phi_0$ and non-RWA error channels).  This compares well with gates designed for similar systems using state-of-the-art numerical optimal control methods.  For example, Abdelhafez \textit{et al.}~\cite{Mohamed2020Universal} used numerical optimal control to design gates in a fluxonium circuit operated in a $T_1$-protected regime.  They constrained the drive power in a manner similar to our approach, but used slightly different fluxonium parameters and a lower level of flux noise  ($A_{\Phi_{\mathrm{ext}}} = 1\mu\Phi_0$, $3$ times smaller than in our work).
They achieved an X gate with $\bar{F} \approx 0.996$ in time $t_g = 60$ ns.  The main difference in parameters is that Abdelhafez \textit{et al.}~\cite{Mohamed2020Universal} used flux bias $\Phi_{\mathrm{ext}} = 0.45\Phi_0$ (versus $\Phi_{\mathrm{ext}} = 0.17\Phi_0$ in our work), and had more isolated qubit states: $|n_{01}| = 0.01$ in their work, a factor of two smaller than in our model system.

Our results thus show that even the inclusion of the effects of $1/f$ dephasing, our SATD tripod gate can achieve excellent performance. These results are essentially unchanged if one now also includes $T_1$ dissipation, as long as the relevant $T_1$ times are long enough.  Appendix~\ref{sec:ErrorT1} provides a thorough analysis of our gate performance obtained by our including realistic $T_1$ effects due to dielectric losses (which is believed to be the dominant mechanism in fluxonium~\cite{Nguyen2019High,Helin2021Universal}).  We find that high gate fidelities are still possible:  for example, the gate fidelity at $t_g = $ 100 ns is 0.9991-0.9997 for reasonably high but realistic~\cite{Wang2019Cavity,smith2020superconducting} values of dielectric quality factor $Q_{\mathrm{diel}} \geq 10^6$.

\section{Conclusions}\label{sec:conclusions}
We have presented a general strategy showing how analytic shortcuts-to-adiabaticity techniques can be used even in complex multilevel systems where the rotating-wave approximation is not valid.  Our strategy to mitigate non-RWA errors involves first exploiting the degeneracy of perfect STA protocols in the RWA limit, and then correcting pulses with (analytically derived) frequency chirps.  As a demonstration of our technique, we theoretically analyzed the implementation of an accelerated adiabatic ``tripod'' gate in a realistic multilevel superconducting circuit (a driven fluxonium qubit).  We focused on parameter regimes where the qubit levels are highly isolated, yielding $T_1$ protection but making traditional gates more problematic. Our analysis revealed that our techniques combined with a judicious choice of system parameters yield an accelerated adiabatic gate having competitive performance.  Including realistic levels of $1/f$ flux noise dephasing and $T_1$ dissipation noise, we achieve a gate fidelity of 0.9991-0.9997 in a gate time of 100 ns. We also showed that our approach can compare  favorably against a more straightforward direct-driving approach to gates in this system.  Using power constraints arising from a realistic setup as if the qubit is driven through a cavity, we found that for an X gate,  our approach yields errors more than $5$ times smaller than the direct driving approach.

While our test system consists of a fluxonium qubit, our discussion and methods are very general and can be readily applied to various other architectures. This, along with the fact that the pulses are described analytically, while still providing good performance, will hopefully prove to be very useful in various quantum-control-related applications.   

\acknowledgments 
This work is supported by the Army Research Office under Grant No. W911NF-19-1-0328. We thank Helin Zhang, Srivatsan Chakram, Brian Baker and Jens Koch for fruitful discussions. F.S. is grateful to Long B. Nguyen for enlightening discussions. We acknowledge the University of Chicago Research
Computing Center for support of this work.

\appendix
\section{Basic working of the tripod gate}\label{sec:tripodgate}
In this section, we give a brief overview of the basic four-level geometric tripod gate introduced in Refs.~\cite{duan2001geometric,Kis2002Qubit} [see Fig.~\ref{fig:tripod}(a)]; the discussion here follows Ref.~\cite{Ribeiro2019Accelerated}. The tripod Hamiltonian $\Hstirap(t)$ [Eq.~\eqref{eq:Hstirap}] has two instantaneous zero-energy dark states that span the dark-state manifold and are orthogonal to $|\rme\rangle$.  The basic tripod gate uses the geometric evolution of states in the dark-state manifold. In this manifold, there is always one (time-independent) state (defined by the time-independent control pulse parameters $\alpha$ and $\beta$) that is purely qubitlike:  
\begin{equation}\label{eq:logicalzero}
|\tilde{0}\rangle = \sin(\alpha)|0\rangle - \mathrm{exp}(i\beta)\cos(\alpha)|1\rangle.
\end{equation}
The qubit state orthogonal to this dark state is
\begin{equation}\label{eq:logicalone}
|\tilde{1}\rangle  = \cos(\alpha)|0\rangle + \mathrm{exp}(i\beta)\sin(\alpha)|1\rangle.
\end{equation}
Expressing $\Hstirap(t)$ in these new qubit basis states, we have
\begin{equation}\label{eq:Hlambda}
\Hstirap(t) = \frac{1}{2}\left[\Omega_{\tilde{1}\rme}(t)|\tilde{1}\rangle \langle \rme| + \Omegaae(t) |\rma\rangle \langle \rme| + \mathrm{H.c.} \right],
\end{equation}
where $\Omega_{\tilde{1}\rme} = \Omega_0 \sin[\theta(t)]$ and $\Omega_{\mathrm{ae}} = \Omega_0 \cos[\theta(t)]e^{i\gamma(t)}$ [Eq.~\eqref{eq:omegaae}]. 
In this new basis, the qubit state $|\tilde{1}\rangle$, together with states $|\rma\rangle$ and $|\rme\rangle$, form a three-level $\Lambda$ system~\cite{Bergmann1998Coherent,Vitanov2017Stimulated} [see Fig.~\ref{fig:tripod}(b)]. One can write a geometric phase ~\cite{pancharatnam1956generalized,berry1984quantal} onto the state $|\tilde{1}\rangle$ by performing the ``double STIRAP protocol", where one slowly varies control pulses to realize the cyclic adiabatic  evolution  $|\tilde{1}\rangle \rightarrow |\rma\rangle \rightarrow |\tilde{1}\rangle$.     
The resulting phase is the basis of the tripod adiabatic single-qubit gate~\cite{duan2001geometric,Kis2002Qubit}.   
One can perform an arbitrary single qubit gate in this manner, without requiring precise pulse timing and without requiring direct couplings between the logical qubit states. 

To understand the above double STIRAP protocol in more detail, note that the dark state relevant to our $\Lambda$ system (and orthogonal to $|\tilde{0}\rangle$) is
\begin{equation}
|\rmd(t)\rangle = \cos[\theta(t)]|\tilde{1}\rangle - e^{i\gamma(t)}\sin[\theta(t)]|\rma\rangle.
\end{equation}
The required cyclic adiabatic evolution is achieved by varying the pulse parameter $\theta(t)$, which brings the dark state $|\rmd(t)\rangle$ from the state $|\tilde{1}\rangle$ at $t=0$ to $|\mathrm{a}\rangle$ at $t= t_g/2$ and back to $|\tilde{1}\rangle$ at the final gate time $t = t_g$. To do this, we use a symmetric form for $\theta(t)$, i.e.,~
\begin{equation}\label{eq:theta}
\theta(t) = \begin{cases}
\displaystyle\frac{\pi}{2} P(t/t_g) & \displaystyle 0 \leq t \leq \frac{t_g}{2}\\[8pt]
\displaystyle\frac{\pi}{2} \left[1- P\left(\frac{t}{t_g} - \frac{1}{2} \right) \right] & \displaystyle \frac{t_g}{2}< t \leq t_g,
\end{cases}
\end{equation}
where $P(x)$ is a function that increases monotonically from $P(0) = 0$
 to $P(1/2) = 1$.
Furthermore, to ensure a smooth turn on and turn off of the control fields, we choose a polynomial that gives $\dot{\theta} (0) = \dot{\theta}(t_g/2) = \dot{\theta}(t_g) = \ddot{\theta} (0) = \ddot{\theta}(t_g/2) = \ddot{\theta}(t_g) = 0$. In particular, we use the simplest polynomial satisfying the above criteria which is given by~\cite{Ribeiro2019Accelerated}
\begin{equation}\label{eq:Px}
P(x) = 6 \left(2x\right)^5 - 15 \left(2x \right)^{4} + 10 \left(2x\right)^3.
\end{equation} 
One also needs a nontrivial relative pulse phase $\gamma(t)$ to obtain a net Berry phase.  Following Ref.~\cite{Ribeiro2019Accelerated}, we use the simple form of $\gamma(t)$ as given  in Eq.~\eqref{eq:gamma}:
\begin{equation}
\gamma(t) = \gamma_0 \Theta \left(t - \frac{t_g}{2} \right),
\end{equation}
where $\Theta(t)$ is the Heaviside step function. 

In the adiabatic limit $\dot{\theta}(t)/\Omega_0 \rightarrow 0$, one can show~\cite{Ribeiro2019Accelerated} that the dark state $|\rmd(t)\rangle$  accumulates a geometric phase $\gamma_0$ at $t = t_g$, and the qubit subspace evolves independently of the auxiliary-level subspace.  The net result is a geometric single-qubit gate controlled by the pulse parameters $\alpha$, $\beta$, and $\gamma_0$.  It is  described by the unitary in the qubit subspace $\hat{U}_{\mathrm{G},01}$ given in Eq.~\eqref{eq:UQubitAdiabatic}. The full adiabatic limit unitary has the form
$\hat{U}_{\mathrm{G}} = \hat{U}_{\mathrm{G},01} \oplus \hat{U}_{\mathrm{G,ae}}$ where $\hat{U}_{\mathrm{G,01}}$ and $\hat{U}_{\mathrm{G,ae}}$ are the unitaries acting in the qubit and auxilary subspaces, respectively (see Ref.~\cite{Ribeiro2019Accelerated}). 

\section{SATD dressing for STA protocols}\label{sec:SATD}
We review briefly how the ``dressed state" approach to constructing STA protocols \cite{Baksic2016Speeding,Ribeiro2019Accelerated} can be used to accelerate the tripod gate \cite{Ribeiro2019Accelerated}.  The general goal is to have the system follow a ``dressed" version of the original adiabatic eigenstate that coincides with the original state at the start and end of the protocol.  This can be achieved by using a time-dependent dressing function $\nu(t)$ that vanishes at $t = 0$ and $t = t_g$. Following Refs.~\cite{Baksic2016Speeding,Ribeiro2019Accelerated}, we introduce $|\rmd_{\nu}\rangle$, a dressed version of the original dark state $|\rmd\rangle$ [Eq.~\eqref{eq:d2}]: 
\begin{equation}\label{eq:dressingJ}
    |\rmd_\nu(t)\rangle = \exp \left[-i\nu(t)\hat{J}_{x}\right] |\rmd(t)\rangle.
\end{equation}
Here $\hat{J}_x = [|\mathrm{b}_+(t)\rangle \langle  \rmd(t)| + |\mathrm{b}_-(t)\rangle \langle \rmd(t) | +\mathrm{H.c.}]/\sqrt{2}$, and $|\mathrm{b}_\pm(t)\rangle$ denote the bright adiabatic eigenstates of $\Hstirap(t)$ with energies $\pm \Omega_0/2$.  

As discussed in Ref.~\cite{Ribeiro2019Accelerated}, for the phase accumulated by this state to be purely geometric and equal to the adiabatic-limit geometric phase $\gamma_0$, we require $\nu(t_g/2) =0$. A particular dressing that satisfies this constraint is SATD~\cite{Baksic2016Speeding,Ribeiro2019Accelerated}, where the dressing angle $\nu$ is given by
\begin{equation}\label{eq:nusatd}
\nu(t) = \nu_{\mathrm{SATD}}(t) \equiv \arctan \left[\frac{2\dot{\theta}(t)}{\Omega_0} \right].
\end{equation}
Using the SATD dressing function, one can show~\cite{Baksic2016Speeding,Ribeiro2019Accelerated} that the accelerated protocol is implemented by modifying the original uncorrected pulse according to Eq.~\eqref{eq:omegaSATD}.

\section{Comparing SATD against simple nonadiabatic protocols}
\label{sec:lowerbound}

In Sec.~\ref{subsec:RMSvoltage}, we discussed how $\Omegarms$, the time-averaged adiabatic gap of our Hamiltonian, is a relevant metric of our protocol's energy cost.  We also discussed that by optimizing SATD, one can achieve an energy cost $\Omegarms/2\pi =  1.92/ t_g$, where $t_g$ is the gate time.  In this appendix, we show that this compares surprisingly favorably with simpler, nonadiabatic pulse protocols. 

First, consider the double-swap and hybrid scheme protocols (see, e.g., Refs.~\cite{wang2012using,Wang2012using0}).  To understand these protocols, consider the effective three-level $\Lambda$ system as shown in Fig.~\ref{fig:tripod}(b) [with the  Hamiltonian given in Eq.~\eqref{eq:Hlambda}].  The double-swap protocol involves two sequential swap operations for each STIRAP process, where in the first half of the protocol the pulse
$\Omega_{\tilde{1}\mathrm{e}}(t)$ 
is first turned on with a constant value of $\Omega_0$ for half of the time and is then turned off with a simultaneous turn on of the pulse $\Omega_{\mathrm{ae}}(t)$  with a constant value of $\Omega_0$ for the other half of the time. The whole sequence is then reversed for the second half of the protocol. It follows that to generate a geometric quantum gate tjat cyclically evolves the state $|\tilde{1}\rangle \rightarrow |\rme\rangle \rightarrow |\rma\rangle \rightarrow |\rme\rangle \rightarrow  |\tilde{1}\rangle$,  the protocol must be executed for a total gate time  $t_g = 2\pi/(\Omega_0/2)$. So, we have $\Omegarms = \Omega_0 = 4\pi/t_g$.  As shown in Fig.~\ref{fig:Optimalomegagate}, this is slightly larger than the energy cost of the optimized SATD protocol.

Alternatively, consider the hybrid scheme.  This has both pulses [$\Omega_{\tilde{1}\mathrm{e}}(t)$ and $\Omega_{\mathrm{ae}}(t)$] turned on for the whole protocol with a constant value of $\Omega_0$. For a cyclic evolution of the state, the protocol must then be performed for a total gate time $t_g = 2\pi/(\sqrt{2}\Omega_0/2)$. Since both pulses are turned on for the whole protocol, we have $\Omegarms = \sqrt{2}\Omega_0 = 4\pi/t_g$, which is the same as for the double-swap protocol. 

\subsection{Proof for the lower bound of $\Omegarms$}
Finally, we establish a rigorous lower bound on $\Omegarms$ using the quantum speed limit of Ref.~\cite{Pires2016Generalized}, which generalized previous work~\cite{mandelstam1991uncertainty,margolus1998maximum}. The lower bound on $\Omegarms$ that we derive below holds for any generic dressing function $\nu(t)$ [Eq.~\eqref{eq:dressingJ}], including the SATD dressing function [Eq.~\eqref{eq:nusatd}]. We begin by applying the bound in Ref.~\cite{Pires2016Generalized} to the first half of our gate protocol.  Letting 
$\hat{\rho}_0$ ($\hat{\rho}_{t_g/2}$) denote the initial system state (state after evolution for a time $t_g/2$), we have
\begin{equation}\label{eq:QSL}
    \LQF(\hat{\rho}_0,\hat{\rho}_{t_g/2}) \leq \frac{1}{\hbar}\int_{0}^{
    t_g/2}dt \sqrt{\langle \Hzero^2(t)\rangle  - \langle \Hzero(t) \rangle^2},
\end{equation}
where $\LQF(\rho_0,\rho_{t_g/2}) = \arccos\left[\sqrt{F(\rho_0,\rho_{t_g/2})}\right]$ where
$F(\rho_0,\rho_{t_g/2}) = \mathrm{Tr}\left(\sqrt{\sqrt{\rho_0}\rho_{t_g/2}\sqrt{\rho_0}}\right)$ is the Uhlmann fidelity.  The left-hand side is the distance of the initial and final states according to the quantum Fisher information metric.  The right-hand side of this inequality is the time-integrated instantaneous energy uncertainty of our Hamiltonian $\Hzero(t)$ with the accelerated protocol.  The symmetry of our protocol implies the left-hand side also bounds the energy uncertainty over the interval $(t_g/2, t_g)$. 

To apply this to our accelerated protocol, note that during the first half of the evolution, the zero-energy dark state $|\rmd(t)\rangle\langle\rmd(t)|$ [Eq.~\eqref{eq:d2}] evolves from the initial state $\rho_0 =|\tilde{1}\rangle\langle\tilde{1}|$ to an orthogonal  state $\rho_{t_g/2} = |\mathrm{a}\rangle \langle \mathrm{a}|$.  As a result, the LHS of Eq.~\ref{eq:QSL} becomes $\pi/2$.

For a generic dressing function $\nu(t)$, we can obtain the accelerated Hamiltonian $\Hzero(t)$ in the right-hand side of Eq.~\eqref{eq:QSL} from the adiabatic Hamiltonian [Eq.~\eqref{eq:Hlambda}] by modifying the original pulse angle and amplitude via~\cite{Baksic2016Speeding} 
\begin{subequations}\label{eq:transformation}
\begin{align}
\theta(t) &\rightarrow \tilde{\theta}(t) = \theta(t) +\arctan\left(\frac{\dot{\nu}(t)}{\dot{\theta}(t)/\tan[\nu(t)]} \right), \\
\Omega_0 &\rightarrow \tilde{\Omega}(t) = \sqrt{\dot{\nu}^2(t)+\left(\frac{\dot{\theta}(t)}{\tan [\nu(t)]} \right)^2}\label{eq:omega0transform}.
\end{align}
\end{subequations}
Using the fact that the accelerated protocol guarantees that the system's state follows (at all times) the ``dressed'' dark state $|\mathrm{d}_\nu(t)\rangle$ defined in Eq.~\eqref{eq:dressingJ}, we can  calculate the instantaneous energy uncertainty (right-hand side of Eq.~\eqref{eq:QSL}) from the probability $p_{\pm}$ of the dressed dark state being in the instantaneous eigenstates $|\rmbpm(t)\rangle$ of $\Hzero(t)$. Here $|\rmbpm(t)\rangle$ are the bright states of $\Hzero(t)$ with eigenenergies $\pm \tilde{\Omega}/2$:
\begin{equation}
|\rmbpm(t)\rangle = \frac{1}{\sqrt{2}}\left(\pm \sin[\tilde{\theta}(t)]|\tilde{1}\rangle \pm e^{i\gamma(t)}\cos[\tilde{\theta}(t)]|\rma\rangle + |\rme\rangle \right).
\end{equation}
The probabilities $p_{\pm}$ are given by
\begin{align}\label{eq:population}
p_{\pm}(t) &= \left|\left \langle \mathrm{d}_{\nu}(t) |\rmbpm(t) \right\rangle\right|^2 \nonumber\\
&= \frac{1}{2}\left[\left( \frac{\dot{\nu}^2(t)}{\tilde{\Omega}^2(t)}\right) \cos^2\left[\nu(t)\right]+ \sin^2\left[\nu(t)\right] \right].
\end{align}
Using Eq.~\eqref{eq:population}, we  calculate the instantaneous values of $\langle \Hzero (t)\rangle$ and $\langle \Hzero^2(t)\rangle$ as
\begin{subequations}\label{eq:expH}
\begin{align}
\langle \Hzero (t)\rangle &= \frac{\tilde{\Omega}(t)}{2}(p_+ -p_-) = 0,\\
\langle \Hzero^2(t)\rangle  
&= \frac{[\tilde{\Omega}(t)]^2}{4}(p_+ + p_-) =\frac{1}{4}\left(\dot{\nu}^2(t) + \frac{\dot{\theta}^2(t)}{\sec^2[\nu(t)]}\right)\label{eq:H02}.
\end{align}
\end{subequations}
Substituting Eq.~\eqref{eq:expH} into Eq.~\eqref{eq:QSL}, we then have
\begin{align}\label{eq:QSLineq}
\frac{\pi}{2} &\leq \frac{1}{2}\int_{0}^{t_g/2} dt \sqrt{ \dot{\nu}^2(t) + \frac{\dot{\theta}^2(t)}{\sec^2[\nu(t)]}}\nonumber\\
 &\leq \frac{1}{2}\sqrt{\int_{0}^{t_g/2} dt}\sqrt{ \int_{0}^{t_g/2} dt \left(\dot{\nu}^2(t) + \frac{\dot{\theta}^2(t)}{1+\tan^2[\nu(t)]}\right)}\nonumber\\
 & < \frac{t_g}{4}\sqrt{\frac{2}{t_g} \int_{0}^{t_g/2} dt \left(\dot{\nu}^2(t) + \frac{\dot{\theta}^2(t)}{\tan^2[\nu(t)]}\right)}= \frac{t_g}{4} \tilde{\Omega}_{\mathrm{RMS}}.
\end{align}
In  going to the second line of Eq.~\eqref{eq:QSLineq}, we used the Cauchy-Schwarz inequality and the relation $\sec^2(x) = 1+ \tan^2(x) $. From Eq.~\eqref{eq:QSLineq}, we can write the bound for $\Omegarms$ as
\begin{align}
 \tilde{\Omega}_{\mathrm{RMS}}> 2\pi/t_g.
\end{align}

\section{Optimal fluxonium parameter regime for a tripod gate in the $T_1$-protected regime}\label{sec:justification}

To get a tripod gate with small non-RWA errors as well as a qubit with a long $T_{1}$ coherence time, we use the following criteria in choosing the fluxonium circuit parameters:
\begin{enumerate}
    \item The ground states (i.e., low-lying energy levels of the tripod that are labeled by $|0\rangle$, $|1\rangle$, and $|\mathrm{a}\rangle$) should be well isolated from each other and be nondegenerate. The requirement of strong isolation is necessary to obtain a $T_1$-protected qubit. On the other hand, the nondegeneracy of ground states is required to ensure sufficiently large detuning of the spurious crosstalk transitions from the driving frequencies to reduce the coherent errors due to crosstalk. 
\item The charge matrix elements coupling the excited state to the ground states of the tripod system should be large and have the same order of magnitude. This requirement helps to minimize coherent errors arising from non-RWA processes. 
\end{enumerate}

Criterion (1) requires us to pick circuit parameters that satisfy $E_J \gg E_C$ and $E_{L} \ll E_{J}$ (for well-localized ground states) as well as $0<\Phi_{\mathrm{ext}} \lesssim \Phi_0/4$ (to lift the degeneracy of the ground states). Criterion (2), on the other hand, requires that we pick the excited state $|\mathrm{e}\rangle$ of the tripod gate to be the first excited state of the central well that is delocalized over the potential wells where the ground states $\ket{0}$, $\ket{1}$, and $\ket{\mathrm{a}}$ reside but be somewhat separated from the much more densely-spaced higher energy levels. This means that the state $\ket{\mathrm{e}}$ must lie in the vicinity of the top edge of the cosine potential, requiring $\sqrt{8 E_{C} E_{J}} \gg 2 E_{J}$. It is clear that criteria 1 and 2 cannot be satisfied simultaneously in a standard fluxonium circuit. That forces us to seek a balanced parameter set, ensuring that we can end up with both a long-lived qubit and a tripod that allows a high-fidelity SATD gate.

To achieve this balance, we initially do a numerical search over experimentally realizable circuit parameters, enforcing conditions on the energy-level structure that do not strongly violate our desired selection criteria outlined in criteria 1 and 2 above. These include, for example, requiring that the charge matrix element between levels $\ket{0}$ and $\ket{1}$ be small (which maximizes $T_{1}$), and that the tripod transition energies $\omega_{j\mathrm{e}}$, with $j=0,1,\mathrm{a}$, are not degenerate (to minimize effects of crosstalk). 
After applying this procedure, we end up with a much reduced parameter space, which in turn is used to perform a more focused search on the circuit parameter set that optimizes the fidelity of an actual SATD gate (although without the frequency chirping). 
Taking all of the above factors into account, we end up with a final choice of parameters: $E_L/h = 0.063$ GHz, $E_J/h = 9.19$ GHz, $E_C/h = 2$ GHz, and $\Phi_{\mathrm{ext}} = 0.17 \Phi_0$. We stress that the small inductive energy of our fluxonium puts our $E_L$ in same regime as the Blochnium device that was recently realized experimentally~\cite{pechenezhskiy2020superconducting}. The corresponding energy level structure and the potential energy landscape are shown in Fig.~\ref{fig:tripodfluxonium}. 
Because of the small matrix element between qubit levels $\ket{0}$ and $\ket{1}$ ($ |n_{01}|= 0.02$), our qubit is not $T_{1}$ limited, and because of the positioning of the excited level $\ket{\mathrm{e}}$, a tripod with relatively strong tripod matrix elements ($|n_{0\mathrm{e}}|= 0.27$, $|n_{1\mathrm{e}}|= 0.46$, and $|n_{\mathrm{a}\mathrm{e}}|= 0.16$)  can be realized. 

\section{Accounting for a smooth turn on and turn off of the pulse at the beginning and end of the protocol}\label{sec:ramp}
 Since realistic pulses are off at the beginning and the end of the protocol, we sandwich the pulse in Eq.~\eqref{eq:omegadrive} by a ramp time $\tramp$ during which the pulse $\Omegaaetilde(t)$ is smoothly turned on (off)  at the beginning (end) of the protocol. The full driving pulses can then be written as smooth piecewise continuous functions that can be separated into three time regions [region (I): $0\leq t < \tramp$; region (II): $\tramp \leq t \leq t_g + \tramp$; region (III): $ t_g +\tramp < t \leq t_g + 2\tramp$)] as
\begin{subequations}\label{eq:realisticpulse}
\begin{align}
\frac{\Omegazeroetilde (t)}{\Omega_0 } &= \begin{cases}
0 & \rm{(I)},\\
\displaystyle\cos\alpha\left\{\sin[\theta(t_-)] + \frac{\cos[\theta(t_-)]\ddot{\theta}(t_-)}{ \dot{\theta}^2(t_-)+\Omega_0^2 /4}\right\}  & \rm{(II)},\\
 0 & \rm{(III)}, 
\end{cases}\\
\frac{\Omegaoneetilde (t)}{\Omega_0 } &= \begin{cases}
0 & \rm{(I)},\\
\displaystyle e^{-i\beta}\sin\alpha \left\{\sin[\theta(t_-)] +  \frac{\cos[\theta(t_-)]\ddot{\theta}(t_-)}{  \dot{\theta}^2(t_-)+\Omega_0^2/4}\right\} & \rm{(II)}, \\
0 & \rm{(III)},
\end{cases}\\
\frac{\Omegaaetilde (t)}{\Omega_0 } &= \begin{cases}
\displaystyle P\left(\frac{t}{2\tramp}\right) & \rm{(I)},\\[10pt]
\displaystyle e^{-i\gamma t_-} \left\{\cos[\theta(t_-)] -  \frac{\sin[\theta(t_-)]\ddot{\theta}(t_-)}{  \dot{\theta}^2(t_-)+\Omega_0^2/4}\right\}& \rm{(II)},\\[10pt]
\displaystyle 1- P\left(\frac{t_- - t_g}{2\tramp}\right) & \rm{(III)},
\end{cases}\label{eq:omegaaeturn}
\end{align}
\end{subequations}
where $t_- = t-t_{\mathrm{ramp}}$.
Specifically, we choose $\tramp = 0.01 t_g$ to be short enough compared with the overall pulse length such that it will not significantly affect the whole dynamics but long enough such that there is no sharp jump in the pulse when it is turned on or off. In Eq.~\eqref{eq:omegaaeturn} we have used the function $P(x)$ given in Eq.~\eqref{eq:Px}  to ensure a smooth turn on and turn off of the pulse $\Omegaaetilde(t)$ during a duration $t_{\mathrm{ramp}}$ at the beginning and end of the protocol. The time profile of the driving pulse with the inclusion of the ramp function is shown in Fig.~\ref{fig:pulse}(a).

\section{Effects of the $T_1$ decay process on the gate error}\label{sec:ErrorT1}
In this section, we discuss  the effects of $T_1$ decay on the gate error. We focus only on the most dominant $T_1$ relaxation process that involves the computational states (i.e., the relaxation from the state $|\mathrm{e}\rangle$ to the state $|\mathrm{1}\rangle$). We have checked that the other $T_1$ relaxation processes affecting the computational states are much longer and hence do not impact the gate performance significantly. We consider the $T_1$ decay process to be due to the  dielectric loss in capacitors since it is typically the most significant relaxation process in experiments~\cite{Nguyen2019High,Helin2021Universal}. The $T_1$ relaxation time for any $|k\rangle\rightarrow|l\rangle$ transition due to the dielectric loss is given by
\begin{equation}\label{eq:T1diel}
1/(T_1)_{kl} = \frac{ |\omega_{lk}|\omega_{lk}}{8E_C Q_{\mathrm{diel}}} \left[\coth\left(\frac{\omega_{lk}}{2k_{\mathrm{B}} T} \right)+1\right] |\langle l|\hat{\varphi}|k\rangle|^2,
\end{equation}
where $Q_{\mathrm{diel}}$ is the dielectric quality factor, $E_C$ is the capacitive energy, $\omega_{lk} \equiv \varepsilon_k - \varepsilon_l$ is the energy transition between the state $|l\rangle$ and the state $|k\rangle$, and $\hat{\varphi}$ is the phase operator. Table~\ref{table:relaxation} shows the $\Toeo$ relaxation time, corresponding to the $|\mathrm{e}\rangle\rightarrow |1\rangle$ decay, calculated at zero temperature $(T=0)$ for different values of $Q_{\mathrm{diel}}$.
We take into account this relaxation process by adding the relaxation Lindblad operator 
\begin{equation}
\hat{Z} = \sqrt{1/\Toeo} |1\rangle\langle e|   
\end{equation}
to the master equation [Eq~\eqref{eq:Lindblad}].

\begin{table}[h]
\begin{tabular}{|c |c |} 
\hline
$Q_{\mathrm{diel}}$ & $\Toeo$ ($\mu$s)\\
\hline
$5\times 10^{5}$& 11.9 \\
$1\times 10^{6}$& 23.8 \\
$2\times 10^{6}$ & 47.6\\
$1\times 10^{7}$ & 238\\
\hline
\end{tabular}
\caption{Relaxation times $\Toeo$ for the $|\mathrm{e}\rangle\rightarrow|1\rangle$ decay. The relaxation times are calculated at zero temperature with use of Eq.~\eqref{eq:T1diel} for different values of the dielectric quality factor $\Qdiel$. For comparison, the qubit dephasing time $T_{\varphi,01} = 7.03$ $\mu$s. The fluxonium parameters used are as for Fig.~\ref{fig:schematic}.
}\label{table:relaxation}
\end{table}

\begin{figure}[t!]
\centering
\includegraphics[width=\linewidth]{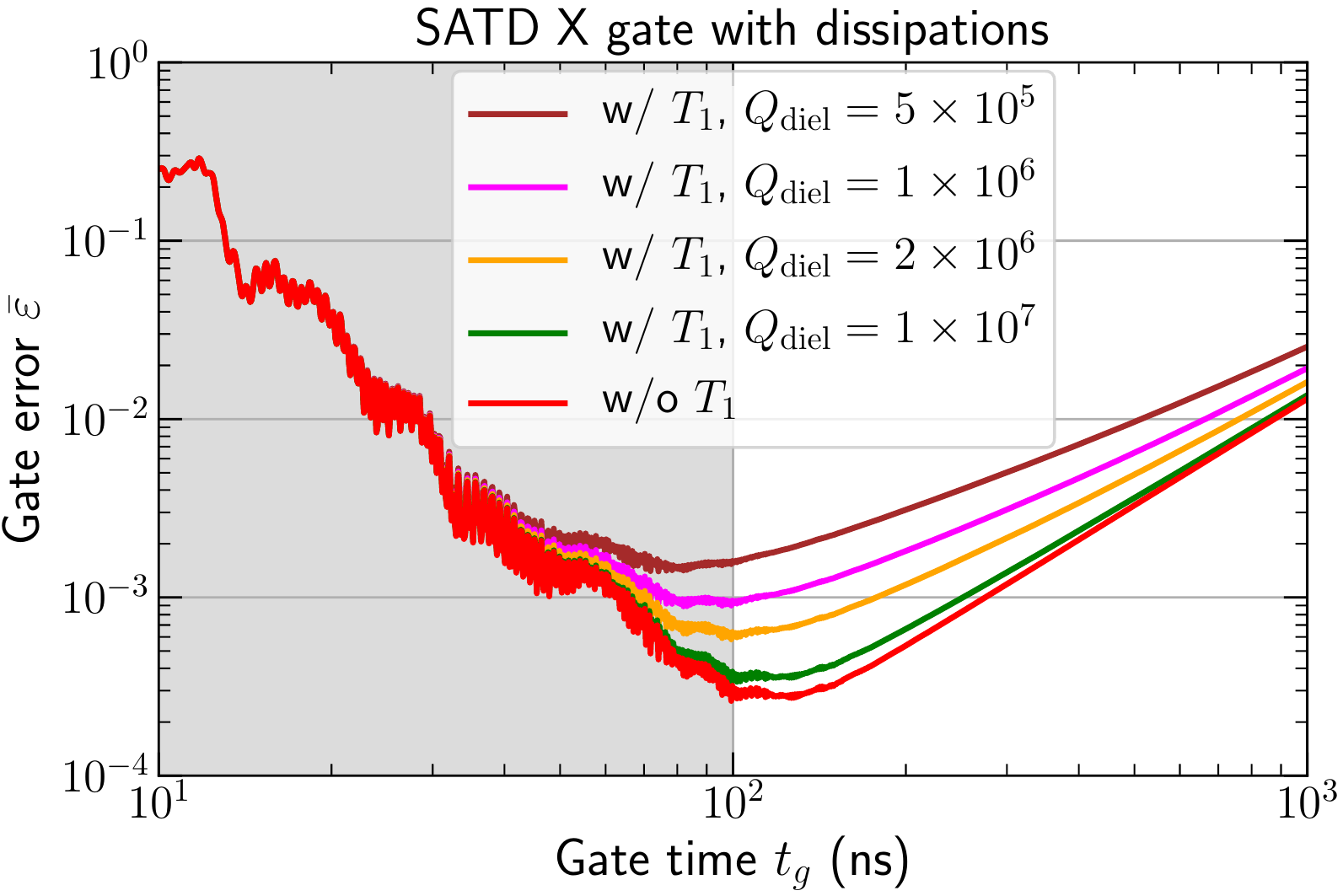}
\caption{
Comparison between the state-averaged X-gate error $\bar{\varepsilon} = 1-\bar{F}$ for SATD protocols with and without $T_1$ relaxation rate. We take into account only the most dominant relaxation process  $\Toeo$, i.e., the relaxation from the state $|\mathrm{e}\rangle$ to the state $|\mathrm{1}\rangle$. The relaxation time $\Toeo$ is calculated with Eq.~\eqref{eq:T1diel} for different values of the dielectric quality factor $\Qdiel$ [see Table~\ref{table:relaxation}  for the  values of $\Toeo$].  The effects of $1/f$ flux noise dephasing as well as non-RWA errors are taken into account; 18 fluxonium levels are included in the simulations. Fluxonium parameters and the dephasing amplitude (corresponding to the qubit dephasing time  $T_{\varphi,01} = 7.03$ $\mu$s) are the same as for Fig.~\ref{fig:Error_dephasecrop}.   
}\label{fig:ErrormapT1}
\end{figure}

Figure~\ref{fig:ErrormapT1} shows a comparison between the state-averaged X-gate error $\bar{\varepsilon}$ for SATD protocols calculated with and without inclusion of the $T_1$ relaxation process. For all plots, we take into account the effects of both $1/f$ flux noise and the coherent errors. As seen in Fig.~\ref{fig:ErrormapT1}, for $(T_1)_{e1}$ values slightly larger than $T_{\varphi,01}$, e.g., $\Qdiel \approx 10^{6}$ [see Table~\ref{table:relaxation} for $(T_1)_{e1}$ values calculated for different $\Qdiel$ values], the $T_1$ relaxation mechanism can still have an appreciable effect on the gate error. This is because our Markovian modeling of the non-Markovian $1/f$ dephasing  noise  (as discussed in Sec~\ref{sec:fluxnoisemodel} and Appendix~\ref{sec:inconsistency}) gives an effective dephasing rate $ t_g/(T_{\varphi,kl})^2$ [Eq.~\eqref{eq:gammak}], which is smaller than the decay rate $1/(T_1)_{kl}$ of a $T_1$ Markovian decay process when $(T_1)_{kl} \approx T_{\varphi,kl}$. Here $T_{\varphi,kl}$ is the pure dephasing time calculated from the standard free-induction-decay calculation for the state $|k\rangle$ with use of the state $|l\rangle$ as the reference state. On the other hand, for high values of $\Qdiel$ (i.e., $\Qdiel \gtrsim 5 \times 10^{6}$), the $T_1$ relaxation process does not significantly change  the gate error.  We note that the use of capacitors with high values of $\Qdiel$ (e.g., $\Qdiel \gtrsim 10^6$) is typical for present experiments~\cite{wang2015surface,smith2020superconducting}. 

\begin{figure}[t!]
\centering
\includegraphics[width=\linewidth]{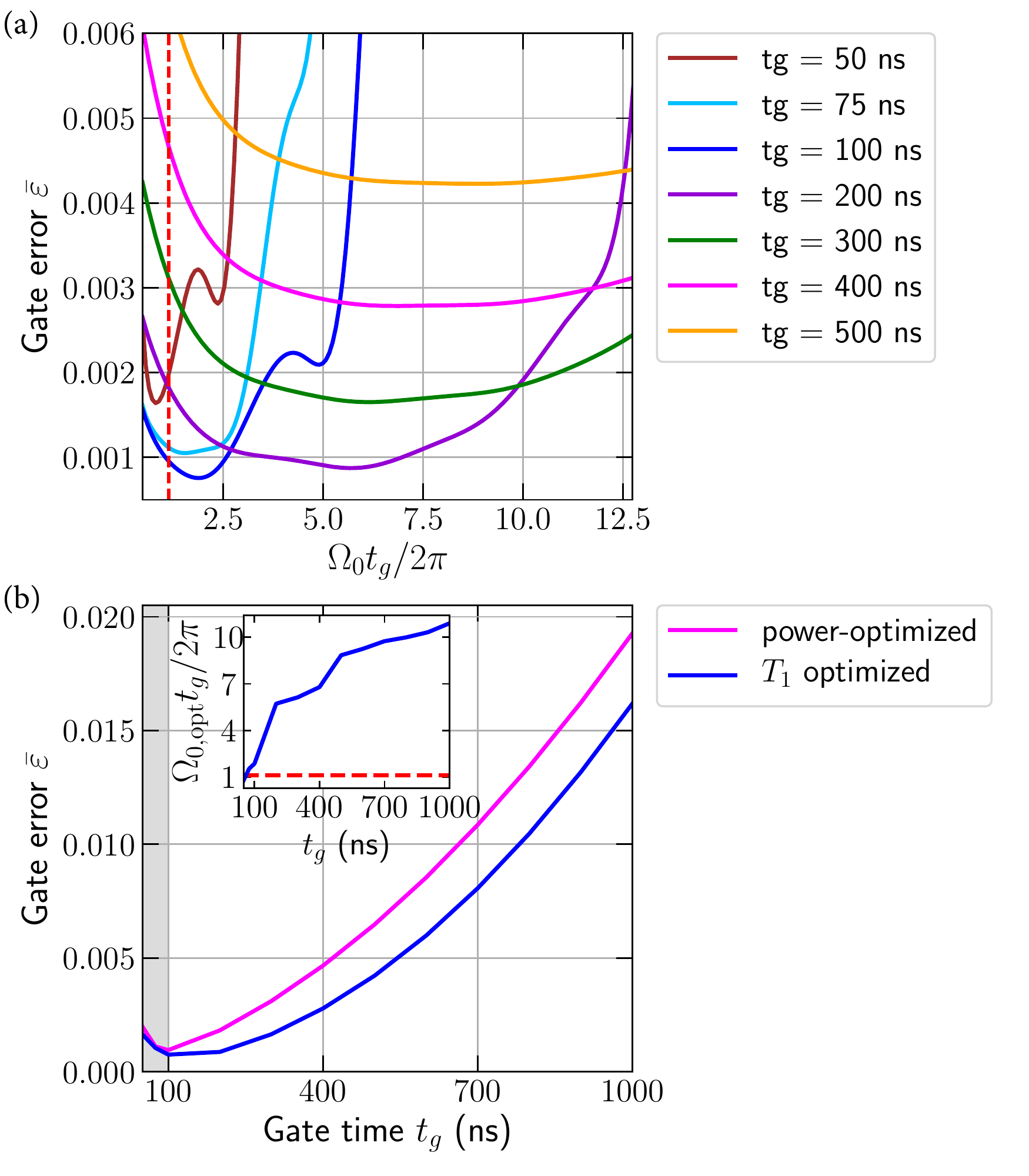}
\caption{
State-averaged SATD X-gate errors $\bar{\varepsilon} = 1-\bar{F}$ calculated in the presence of the $\Toeo$ relaxation process, $1/f$ flux noise dephasing and non-RWA errors. (a) Gate error $\bar{\varepsilon}$ versus $\Omega_0t_g/2\pi$ for different gate times $t_g$. The dashed red line indicates the value of power-optimal $\Omega_0$, i.e., $\Omega_0 t_g/2\pi = 1.135$. (b) Gate error $\bar{\varepsilon}$ versus gate time $t_g$ calculated with the power-optimal $\Omega_0$ (magenta curve) and optimal values of $\Omega_0$ (inset) that minimize the gate errors in the presence of the $T_1$ decay process (blue curve). The shaded (unshaded) region corresponds to the time regime where the gate errors are dominated by non-RWA errors ($T_1$ dissipation). The inset shows the optimal value of $\Omega_0 $ that minimizes the gate error in the presence of the $T_1$ decay process as a function of gate time $t_g$, where the dashed red line corresponds to the power-optimal value  $\Omega_0t_g/2\pi = 1.135$. The relaxation time $\Toeo$ used is $23.8$ $\mu$s which corresponds to dielectric quality factor $\Qdiel = 1\times 10^6$. Eighteen fluxonium levels are included in the simulations. Fluxonium parameters and the dephasing amplitude (corresponding to the qubit dephasing time $T_{\varphi,01} = 7.03$ $\mu$s) are the same as for Fig.~\ref{fig:Error_dephasecrop}.   
}\label{fig:ErrormapT1optimized}
\end{figure}

For cases where the $T_1$ dissipation dominates, we can reoptimize the gate error by using a value of $\Omega_0$  that is larger than the power-optimal $\Omega_0$ [i.e., $\Omega_0 = 1.135(2\pi)/t_g$]. This is because for a fixed $t_g$, SATD protocols populate the excited state $|\mathrm{e}\rangle$ less for larger values of $\Omega_0$~\cite{Baksic2016Speeding}, and hence they are less susceptible to the $T_1$ decay process. However, one cannot use an arbitrarily large value of $\Omega_0$, as beyond a certain value of $\Omega_0$ the coherent errors will start to dominate over the $T_1$ decay process. To this end, we perform a numerical minimization of gate errors in the $T_1$-dominated regime by doing a brute-force optimization of $\Omega_0$. Figure~\ref{fig:ErrormapT1optimized}(a) shows the the gate error versus $\Omega_0t_g$ calculated for different gate times $t_g$. In the coherent-error-dominated regime (i.e., $t_g \lesssim 100$ ns), the value of $\Omega_0$ that gives the minimum gate error is close to the power-optimal $\Omega_0$ [shown as a red dashed line in Fig.~\ref{fig:ErrormapT1optimized}(a) and inset of Fig.~\ref{fig:ErrormapT1optimized}(b)]. However, for the $T_1$-dissipation-dominated regime (i.e., $t_g \gtrsim$ 100 ns), the optimal value of $\Omega_0$ that minimizes the gate error at fixed $t_g$ is larger than the power-optimal $\Omega_0$ (see the inset in Fig.~\ref{fig:ErrormapT1optimized}(b)). In Fig.~\ref{fig:ErrormapT1optimized}(b), we compare the gate errors calculated with the power-optimal $\Omega_0$ (magenta curve) with those obtained with  optimal values of $\Omega_0$ (one for each $t_g$ value) that minimize the gate errors in the presence of the $T_1$ decay process (blue curve). In the coherent-error-dominated regime (shaded region), the minimum gate errors are approximately the same as those obtained  with the power-optimal $\Omega_0$. However, in the $T_1$-decay-dominated regime, there is a reduction in the gate errors obtained with use of the  reoptimized values of $\Omega_0$.
 
\section{Approximating non-Markovian noise with Markovian dynamics}\label{sec:inconsistency}
In this appendix, we show that approximating non-Markovian dynamics of $1/f$ noise with the Markovian master equation cannot capture the decay rate of all coherences correctly.  
In the main text, we constructed our Linblad master equation by picking the decay rate $\Gamma_k$ of the coherences $\rho_{k1} \equiv \langle k|\hat{\rho}|1\rangle$ (coherences that involve the reference level, i.e., the qubit state $|1\rangle$) such that the final decay of the coherences $\rho_{k1}$, for all $ k \neq 1$, at the gate time $t_g$ is  the same for our Markovian dynamics as it would be following the non-Markovian, Gaussian-lineshape decay with an envelope of the form $\exp \left[-(t/T_{\varphi,kl})^2\right]$, where $T_{\varphi,kl}$ is the free induction decay time given by Eq.~\eqref{eq:dephasingrate}. (Note that we can equivalently choose the qubit state $|0\rangle$ as the reference level as both choices will result in the same value of state-averaged fidelities.) While the decay of $\rho_{k1}$ coherences can be captured correctly, the dephasing times of coherences that do not involve the reference level ($\rho_{kl}$, for all $k,l \neq 1$) in our Markovian dynamics are in general not the same as the free induction decay times calculated from Eq.~\eqref{eq:dephasingrate}. To see this, we can write the Lindblad master equation [Eq.~\eqref{eq:Lindblad}] in terms of the density matrix elements as 
\begin{align}\label{eq:Lindbladelem}
\dot{\rho}_{kl}(t) &= -i [\hat{H}(t), \hat{\rho}(t)]_{kl}\nonumber\\
&\hspace{0.4cm}- \left(\mathrm{sgn}\left(\frac{\partial \varepsilon_k}{\partial \Phi_{\mathrm{ext}}} \right)\sqrt{\Gamma_{k}}- \mathrm{sgn}\left(\frac{\partial \varepsilon_l}{\partial \Phi_{\mathrm{ext}}} \right)\sqrt{\Gamma_{l}}  \right)^2\rho_{kl}\nonumber\\
&=  -i [\hat{H}(t), \hat{\rho}(t)]_{kl} - \frac{t_g}{\left(T^{\mathrm{eff}}_{\varphi,kl}\right)^2} \rho_{kl},
\end{align}
where we have identified
\begin{align}\label{eq:Lindbladtime}
\frac{1}{T^{\mathrm{eff}}_{\varphi,kl}} \equiv \left|\mathrm{sgn}\left(\frac{\partial \varepsilon_k}{\partial \Phi_{\mathrm{ext}}} \right)\frac{1}{T_{\varphi,k1}} - \mathrm{sgn}\left(\frac{\partial \varepsilon_l}{\partial \Phi_{\mathrm{ext}}} \right) \frac{1}{T_{\varphi,l1}}\right|.  
\end{align}
We can see that the effective dephasing time $T^{\mathrm{eff}}_{\varphi,kl}$ [Eq.~\eqref{eq:Lindbladtime}], for all $k,l \neq 1$  calculated from the Lindblad master equation is in general not the same as the free induction dephasing time $T_{\varphi,kl}$  calculated from Eq.~\eqref{eq:dephasingrate}. As shown in Table~\ref{table:dephasing}, our approach if anything overestimates the dominant dephasing processes within the tripod subspace, e.g., $T^{\mathrm{eff}}_{\varphi,\mathrm{a0}}$ and $T^{\mathrm{eff}}_{\varphi,\mathrm{ae}}$.

\begin{table}[h]
\begin{tabular}{|c |c |c |} 
\hline
$kl$ & $T_{\varphi,kl}$ ($\mu$s)& $T^{\mathrm{eff}}_{\varphi,kl}$ ($\mu$s)\\
\hline
01& 7.03 & 7.03 \\
a1 & 6.97 & 6.97 \\
e1 & 53.43 & 53.43 \\
a0 & 3.50 & 1.75 \\
e0 & 8.09 & 17.31 \\
ae & 6.16 & 3.76 \\
\hline
\end{tabular}
\caption{Dephasing times for transitions between computational states $|k\rangle$ and $|l\rangle$ calculated by two different methods: directly from the free induction decay formula [Eq.~\eqref{eq:dephasingrate}] (middle column) and indirectly from the Lindblad master equation [Eq.~\eqref{eq:Lindbladtime}] (right column). }\label{table:dephasing}
\end{table}

\section{Comparison with Raman gates}\label{sec:Raman}

In this section, we compare the performance of our tripod gate against that of the Raman gate. In the Raman protocol,  detuned drives couple the excited state $|\mathrm{e}\rangle$ to the qubit states $|0\rangle$ and $|1\rangle$, leading to an effective direct coupling between the qubit states~\cite{Vitanov2017Stimulated}. The requirement that the drives in a Raman approach be highly detuned means that (for fixed drive amplitudes) it will result in a much  slower gate than is possible, for example,  by using our SATD tripod approach (which uses resonant drive).

To make the comparison more explicit, let $\Delta$ denote the detunings of both drives used in a Raman protocol and let $\Omega_0$ denote the drive amplitudes.  The Raman gate is based on adiabatically eliminating the excited state to generate an effective qubit-only Hamiltonian.  This elimination necessarily requires  $\Omega_0 \ll \Delta$, leading to a Raman gate time $t_g = 2\pi\Delta/\Omega_0^2 \gg 1/\Omega_0$.  In contrast, using our tripod gate, we are able to achieve much shorter gate times $t_g \sim 1/\Omega_0$ for the same drive amplitude.  As a result, we expect the Raman gate to be more susceptible to dissipation than our tripod gate, which gives an enormous advantage for the tripod gate if dissipation prevents the use of long gate times.   Moreover, unlike the direct-driving and Raman gates, our SATD gate  is robust against imperfections in control pulses~\cite{Ribeiro2019Accelerated}.
 
\section{Indirect qubit driving through a coupled cavity}
\label{sec:resonatorfull}

To provide Purcell protection, in practice, superconducting qubits are often driven indirectly through a coupled cavity. In this appendix, we discuss the details of such driving of our fluxonium qubit, especially in the context where the drive power is constrained such that additional dissipative mechanisms due to the cavity can be avoided. In the following, we show how the driving field applied to the qubit-coupled cavity is related to the driving field seen by the qubit $V(t)$ [Eq.~\eqref{eq:V}]. 
\subsection{Driving field for the cavity}\label{sec:resonator}
In this subsection we derive an explicit relationship between the field that drives the cavity $u(t)$ and the voltage $V(t)$ [Eq.~\eqref{eq:V}] seen by the qubit. We begin by writing the Hamiltonian for the driven cavity-coupled fluxonium as
\begin{align}
\HJC(t) = &\,\sum_{k}\varepsilon_{k}|k\rangle\langle k| + \omegar \hata^\dagger\hata + \sum_{k,l} gn_{kl}|k\rangle \langle l|(\hata^\dagger + \hata)\nonumber\\
& + u(t) (\hata^\dagger + \hata),
\end{align} 
where $\varepsilon_k$ and $|k\rangle$ are the fluxonium eigenenergies and eigenstates, respectively, and $\omegar$ is the cavity frequency. The qubit-cavity coupling strength is $g$, while $n_{kl} = \langle k| \hat{n}|l\rangle$ is the fluxonium charge matrix element. The driving field on the cavity is $u(t)$ and the operators $\hata^\dagger$ and $\hata$ are the cavity photon creation and annihilation operators, respectively. 

We consider operating the cavity-qubit system in the dispersive regime where $g n_{kl} \ll |\varepsilon_{l} - \varepsilon_{k} - \omegar|$, and applying a highly off-resonant drive to limit the cavity photon population. We also consider the drive strength to be weak enough that the cavity photon number is small so to avoid cavity-induced dissipations. In this weak driving power regime, we can estimate the relationship between the applied cavity field $u(t)$ and the driving field on the qubit $V(t)$ by first solving for the classical cavity field independently (i.e., in the limit $g\rightarrow 0$). To this end, we model the cavity as a stand-alone, damped, driven harmonic oscillator. The Heisenberg equation of motion for the cavity-photon annihilation operator $\hata(t)$ can be written as~\cite{Clerk2010Introduction} 
\begin{align}\label{eq:eom}
\dot{\hata}(t) &= i [\HJC(t),\hata(t)]  - \frac{\kappa}{2} \hata(t) -\sqrt{\kappa} \hat{b}_{\mathrm{in}}(t)\nonumber\\
            &= -i\omegar \hata(t) - i u(t) - \frac{\kappa}{2} \hata(t) -\sqrt{\kappa} \hat{b}_{\mathrm{in}}(t).
 \end{align}
Here $\kappa$ is the cavity photon decay rate (due to photon leakage to the bath) and $\hat{b}_{\mathrm{in}}(t)$ is the standard bath annihilation  operator which represents the noise~\cite{Clerk2010Introduction}. The equation of motion for $\hat{a}^\dagger(t)$ can be obtained by taking the Hermitian conjugate of Eq.~\eqref{eq:eom}. 
By treating the cavity field classically and writing the equation of motion in terms of the mean value of the photon field displacement  $x = \langle \hata + \hata^\dagger \rangle /\sqrt{2}$, we have
\begin{equation}\label{eq:EOMdriven}
\ddot{x}(t) = -\left(\omegar^2 + \frac{\kappa^2}{4} \right)x(t) -\kappa \dot{x}(t)- \sqrt{2} \omegar u(t),
\end{equation}
where we have used $\langle \hat{b}_{\mathrm{in}}(t) \rangle = \langle \hat{b}_{\mathrm{in}}^\dagger(t) \rangle = 0$.
The solution of Eq.~\eqref{eq:EOMdriven} is
\begin{equation}
x(t) = x_p(t) + x_h(t),
\end{equation}
where
\begin{equation}
x_h(t) = A e^{-\frac{\kappa}{2}t  }\sin(\omegar t + \phi),
\end{equation}
is the homogeneous solution and $x_p(t)$ is the inhomogeneous solution. To solve for $x_p(t)$, we first define the Fourier transforms of $x(\omega)$ and $u(\omega)$  as
\begin{align}
x(\omega) &= \int_{-\infty}^{\infty} x(t) e^{-i\omega t} dt, \nonumber\\
u(\omega) &= \int_{-\infty}^{\infty} u(t) e^{-i\omega t} dt,
\end{align}
and their inverse Fourier transforms as
\begin{align}
x(t) &= \frac{1}{2\pi}\int_{-\infty}^{\infty} x(\omega) e^{i\omega t} d\omega, \nonumber\\
u(t) &= \frac{1}{2\pi}\int_{-\infty}^{\infty} u(\omega) e^{i\omega t} d\omega.
\end{align}
Taking the Fourier transform of Eq.~\eqref{eq:EOMdriven}, we have
\begin{align}
\left[-\omega^2 + i\kappa\omega + \omegar^2 + \frac{\kappa^2}{4}\right]x(\omega) &= -\sqrt{2} \omegar u(\omega),
\end{align}
which gives
\begin{align}\label{eq:xomega}
x(\omega) = \frac{\sqrt{2}\omegar u(\omega)}{\omega^2 - i\kappa\omega - \left( \omegar^2 + \kappa^2/4\right)}.
\end{align}
We can relate the mean photon displacement $x(t)$ to the qubit driving voltage $V(t)$ by replacing the cavity degree of freedom in the cavity-qubit coupling Hamiltonian by its classical value, i.e.,
\begin{align}\label{eq:Hcouplingexp}
\hat{H}_{\mathrm{c}} & \approx \sum_{kl} g \langle\hata^\dagger + \hata \rangle n_{kl} |k\rangle \langle l|\nonumber\\
 &= \sqrt{2}g x(t)\sum_{kl} n_{kl}|k\rangle \langle l|.
\end{align}
Comparing Eq.~\eqref{eq:Hcouplingexp} with the qubit driving term in the Hamiltonian in Eq.~\eqref{eq:H}, we can identify
 \begin{equation}
      \sqrt{2}gx(t) = V(t), 
 \end{equation}
where in the frequency domain, it can be written as
\begin{align}\label{eq:conditionpulse}
\sqrt{2} g x(\omega) = \int_{-\infty}^{\infty} V(t') e^{-i\omega t'} dt'.
\end{align}
Substituting Eq.~\eqref{eq:xomega} into Eq.~\eqref{eq:conditionpulse}, we have
\begin{align}\label{eq:uomega}
u(\omega) = \frac{[\omega^2 - i\kappa\omega - \left( \omegar^2 + \kappa^2/4\right)]}{2 g \omegar} \int_{-\infty}^{\infty}  V(t')e^{-i\omega t'} dt'.
\end{align}
Finally, taking the inverse Fourier transform, we arrive at an expression for the time-domain form of the cavity pulse $u(t)$ that is required to generate a qubit drive $V(t)$ as
\begin{align}\label{eq:ut}
u(t) &= -\frac{1}{2g\omegar}\left[\frac{d^2 }{dt^2}+\kappa\frac{d}{dt}+ \omegar^2 + \frac{\kappa^2}{4} \right]V(t).
\end{align}

\subsection{Relation between the RMS voltage and the time-averaged cavity photon number}~\label{sec:voltagetophoton}
In this section, we establish the relationship between the RMS voltage $\VRMS$ of the qubit driving field and the time-averaged cavity photon number  $\ncav$, where
\begin{align}
\ncav &= \frac{1}{t_g}\int_0^{t_g}\langle  \hat{a}^\dagger \hat{a} \rangle dt\nonumber\\ &\approx \frac{1}{t_g}\int_0^{t_g}\langle
\hat{a}^\dagger \rangle \langle \hat{a} \rangle dt = \frac{1}{t_g}\int_0^{t_g} dt |\eta(t)|^2.
\end{align}
Here we treat the photon field classically, which allows us to replace $\langle \hat{a}^\dagger\rangle$ and $\langle \hat{a}^\dagger\rangle$ by the classical field amplitudes $\eta^*(t)$ and $\eta(t)$, respectively. 
Using this classical field in Eq.~\eqref{eq:Hcouplingexp}, we can identify the qubit driving voltage as 
\begin{equation}
V(t) = 2 g \mathrm{Re}[\eta(t)].
\end{equation}
The RMS voltage of the qubit driving field is then given by
\begin{align}
\VRMS &= \sqrt{\frac{1}{t_g}\int_{0}^{t_g} |V(t)|^2} \nonumber\\
&=g\sqrt{2\ncav},
\end{align}
where in evaluating the second  line, we used the  relation
\begin{align}\label{eq:etaj}
\frac{1}{t_g}\int_0^{t_g}[\mathrm{Re}(\eta(t))]^2 &= \sum_j\frac{1}{t_g}\int_{0}^{t_g} |\eta_j(t)|^2 \cos^2(\omega_{j}t+\phi_j)\nonumber\\
&\simeq \frac{1}{2t_g}\int_0^{t_g}|\eta(t)|^2 = \frac{\ncav}{2}.
\end{align}
Without loss of generality, in Eq.~\eqref{eq:etaj}, we have written the photon field as a multicomponent field, i.e., $\eta(t) = \sum_j \eta_j e^{-i(\omega_j t + \phi_j)}$. To avoid cavity-induced dissipation mechanisms, it is preferable to have a small cavity photon number, i.e.,  $\ncav = \langle \hat{a}^\dagger \hat{a} \rangle \ll 1$; for our simulations in the main text, we specifically set $\ncav = 0.05$. This constrains the maximum RMS voltage for a fixed cavity-qubit coupling strength $g$ as shown in Eq.~\eqref{eq:VRMSg}. 
Each vertical cut of the plot in Fig.~\ref{fig:Error_dephasecrop}(b) corresponds to different maximum allowed values of $\VRMS$, where the purple vertical lines show explicitly two different fixed values of the cavity-qubit coupling strength $g$ corresponding to two different maximum allowed values of $\VRMS$.

\subsection{Effects of the cavity on the qubit's $T_{1}$ and $T_{2}$ times}
\label{sec:Tonecav}
In this section we briefly outline the effects of  thermal photons in the cavity on the coherence times of the qubit, and show that they do not limit the performance of our gates.  In particular, we consider a regime where the cavity photon decay rate $\kappa \ll g^2/|\Deltajkrpm|$, for all $k,l$ where $\Deltajkrpm = \varepsilon_l - \varepsilon_k\pm \omegar$ is the detuning of the cavity frequency $\omegar$ from the $|k\rangle \leftrightarrow |l\rangle$ energy transition involving the computational levels.   The first noise channel we consider is the Purcell relaxation time, which we denote as $\Tonecav$ and can be calculated as~\cite{Houck2008Controlling}
\begin{align}\label{eq:Tonecav}
    (\Tonecav)_{jk} &= (\Deltajkrpm)^2/(\kappa g^2 |n_{kl}|^2).
\end{align}
Since all the relevant transitions are sufficiently detuned from the cavity frequency, we expect the Purcell relaxation process to be weak. Another relevant noise channel is due to photon shot noise, and leads to pure dephasing. We denote the corresponding time scale as $\Ttwocav$, which can be approximated by~\cite{Bertet2005Dephasing,Rigetti2012Superconducting}
\begin{align}\label{eq:Ttwocav}
\Ttwocav &= 1/(\kappa \bar{n}_{\mathrm{th}}).
\end{align}
We stress that Eq.~\eqref{eq:Ttwocav} is valid in the limit of small cavity photon decay rate relative to all the relevant dispersive shifts ($\kappa \ll g^2/|\Deltajkrpm|$), a regime that we are interested in. Moreover, we also consider operating in the dispersive limit ($g n_{kl} \ll |\Deltajkrpm|$, for all $k,l$) and small cavity-thermal-photon-number regime ($\bar{n}_{\mathrm{th}} \ll 1$) such that the cavity response can be treated independently of the qubit (see Sec.~\ref{sec:resonator}). As an example, we can pick $\omegar/2\pi = 2 $ GHz (chosen to be sufficiently detuned from all of the transition frequencies in our fluxonium [Fig.~\ref{fig:tripodfluxonium}]), $\kappa/2\pi = 10$ kHz, $g/2\pi = 250$ MHz and $\bar{n}_{\mathrm{th}} = 0.05$ (corresponding to a temperature $T \approx 30$ mK). Substituting the above parameters into Eqs.~\eqref{eq:Tonecav} and~\eqref{eq:Ttwocav}, we obtain $(T_{1,\mathrm{cav}})_{01} = 0.15$ s and $T_{2,\mathrm{cav}} = 0.32$ ms.  Since the cavity-induced relaxation time $(T_{1,\mathrm{cav}})_{01}$ and dephasing time $T_{2,\mathrm{cav}}$ are much larger than the dephasing times due to the $1/f$ flux noise (see Table~\ref{table:dephasing}), we can ignore the effect of cavity-thermal-photon-induced dissipation on the gate dynamics.

\bibliography{manuscript}

\end{document}